\documentclass[aps,prb,superscriptaddress,citeautoscript,twocolumn,notitlepage,longbibliography]{revtex4-1}

\usepackage{graphicx}
\usepackage{amsmath}
\usepackage{amssymb}
\usepackage{braket}
\usepackage[usenames]{color}

\newcommand{\bea}{\begin{eqnarray*}}
\newcommand{\eea}{\end{eqnarray*}}
\newcommand{\bne}{\begin{equation*}}
\newcommand{\ede}{\end{equation*}}

\newcommand{\bnen}{\begin{equation}}
\newcommand{\eden}{\end{equation}}
\newcommand{\bean}{\begin{eqnarray}}
\newcommand{\eean}{\end{eqnarray}}
\newcommand{\bnsn}{\begin{subequations}}
\newcommand{\edsn}{\end{subequations}}

\newcommand{\bna}{\begin{array}}
\newcommand{\eda}{\end{array}}
\newcommand{\bnm}{\begin{enumerate}}
\newcommand{\edm}{\end{enumerate}}

\renewcommand{\vec}[1]{\text{\boldmath{$ #1 $}}}

\begin{document}
\title{Parity-to-charge conversion 
for readout of topological Majorana qubits}

\author{G\'abor Sz\'echenyi}
\affiliation{Institute of Physics, 
E\"otv\"os University, 
H-1117 Budapest, Hungary}

\author{Andr\'as P\'alyi}
\email{palyi@mail.bme.hu}
\affiliation{Department of Theoretical Physics 
and
BME-MTA Exotic Quantum Phases "Momentum" Research Group,
Budapest
University of Technology and Economics, 
H-1111 Budapest, Hungary}

\date{\today}

\begin{abstract}
We theoretically study a scheme to distinguish
the two ground states of a one-dimensional 
topological superconductor, which could serve
as a basis for the readout of Majorana qubits.
The scheme is based on parity-to-charge conversion,
i.e., the ground-state parity of the superconductor
is converted to the charge occupation on a tunnel-coupled
auxiliary quantum dot.
We describe how certain error mechanisms
degrade the quality of the
parity-to-charge conversion process.
We consider 
(i) leakage due to a strong readout tunnel pulse, 
(ii) incomplete charge Rabi oscillations due to slow
charge noise, 
and
(iii) charge relaxation due to phonon emission and
absorption.
To describe these effects, we use simple model
Hamiltonians based on the ideal Kitaev chain,
and draw conclusions to generic one-dimensional
topological superconductors wherever possible.
In general, the effects of the error mechanisms
can be minimized by choosing  a smooth shape and
an optimal strength
for the readout tunnel pulse. 
In a case study based on InAs heterostructure
device parameters, we estimate that 
the parity-to-charge conversion error is 
mainly due to slow charge noise for weak tunnel pulses
and leakage for strong tunnel pulses.
\end{abstract}


\maketitle

\tableofcontents


\begin{figure}
\includegraphics[width=1.0\columnwidth]{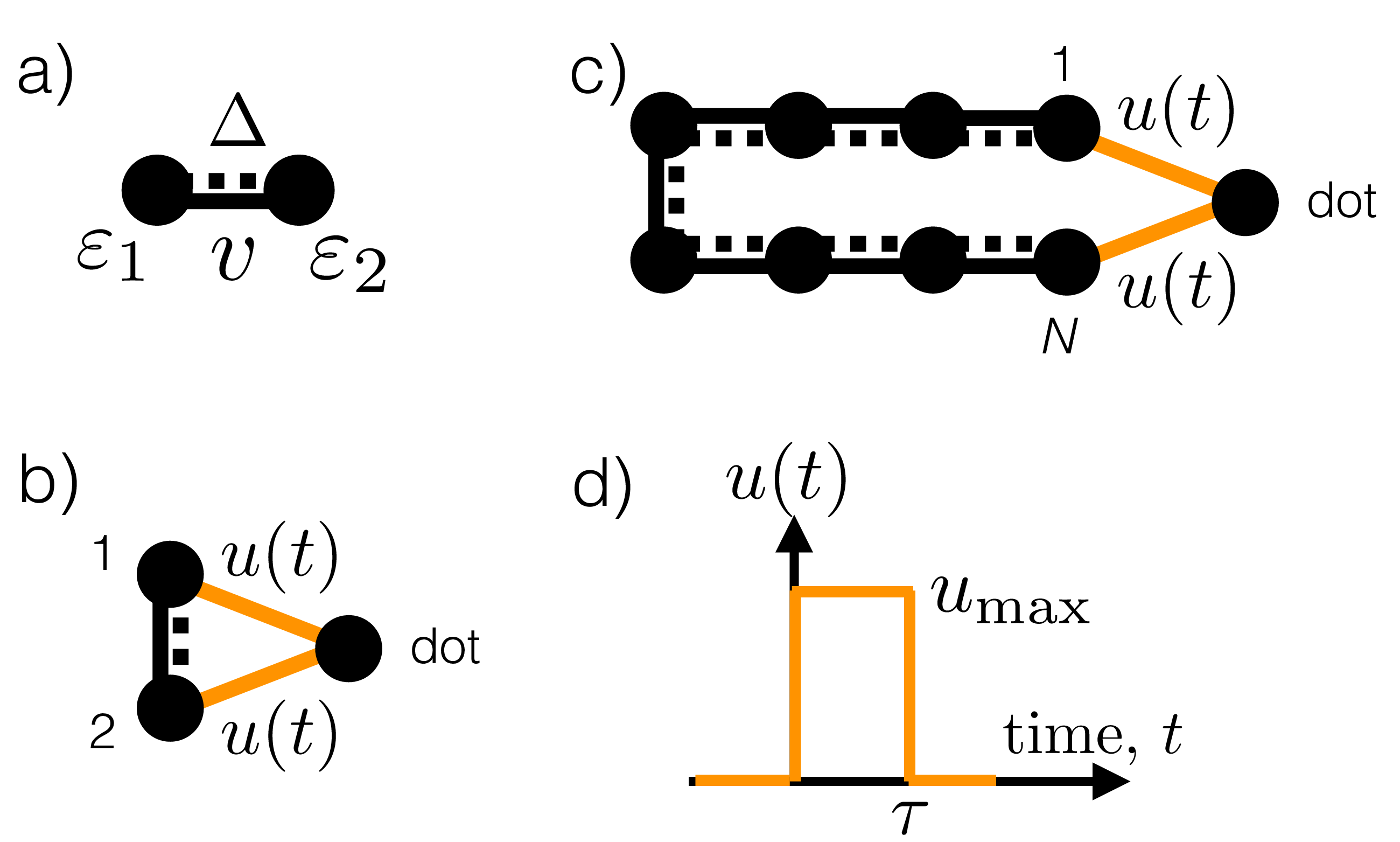}
\caption{
Quantum-dot-assisted parity readout of a 1D topological superconductor.
(a) Two-site Kitaev chain with an effective superconducting gap $\Delta$,
hopping amplitude $v$, and on-site energies $\varepsilon_{1,2}$.
(b) Geometry of a parity readout scheme based on tunneling-induced
charge Rabi oscillation between the chain (1,2) and the readout dot (dot). 
(c) Geometry of the same parity readout scheme with a longer 
Kitaev chain, $N=8$. 
(d) Step-like tunnel pulse inducing the Rabi oscillation between the
chain and the dot. 
}
\label{fig:setup}
\end{figure}

\section{Introduction}

One-dimensional topological superconductors
(1DTSs)
are thought to provide  useful hardware\cite{KitaevChain,Alicea,LeijnseReview,Aasen_milestones}  for 
topological quantum computing\cite{Nayak_rmp}.
The ground state of such a superconductor
is two-fold degenerate, and the two ground states
are distinguished by their fermion parity, 
one being even and the other being odd.
This degeneracy is robust against local perturbations, 
which is a key feature providing protection against
certain types of decoherence.
Recent experiments have been providing more and more
evidence of 1DTS behavior, most notably via transport
experiments 
combining s-wave superconductors with
semiconductors\cite{Mourik,Albrecht,Wiedenmann,Rokhinson,Laroche},
or spinful atomic chains\cite{NadjPergeChainExperiment,HowonKim,Pawlak}.
The natural next step is to try 
quantum information experiments with qubits
based on the robust ground-state degeneracy of
1DTSs, 
often referred to as Majorana qubits \cite{Ramon_Majorana}.

A key primitive toward reading out Majorana qubits is 
the ability to distinguish between the even and 
odd ground states of a single 1DTS, a procedure
that is often called `parity readout'\cite{FlensbergNonabelian,Gharavi,Dmytruk}.
The phrase `parity readout' is also used for the measurement
of the occupation of a single non-local fermionic mode 
shared by two Majorana zero modes\cite{KarzigScalable,Grimsmo,Plugge_2017,Knapp_dephasing,Munk_fidelity,Hyart_majorana}.
The significance of parity readout is especially pronounced
in measurement-only 
topological quantum
computing schemes\cite{Bonderson,KarzigScalable}. 
Direct parity readout 
is hindered by the same features that protect
the Majorana qubit from decoherence.
Therefore, the parity information should first be converted
to a physical quantity that is easy to measure. 
One possibility is to convert the parity information 
to charge information\cite{Gharavi,KarzigScalable}.
We note that in this work, the word `parity' is used
to describe the ground-state fermion number parity, and
not the global fermion number parity, as, e.g., in 
Refs.~\onlinecite{Akhmerov,Hassler}.

Here, we describe a scheme allowing for parity 
readout of a single 1DTS wire (Fig.~\ref{fig:setup}c),
which is a preliminary version of the joint parity readout
of two or four Majorana zero modes. 
In our scheme,  parity readout is achieved via
parity-to-charge conversion and a subsequent 
charge readout. 
The parity-to-charge conversion stage uses
a nearby quantum dot (see Fig.~\ref{fig:setup}b, c), and the parity
information of the wire is converted to the 
occupation of the quantum dot.
Clearly, the error of the parity readout is a key figure of 
merit for the
performance of future devices based on 
Majorana qubits. 
Hence, we consider and quantitatively 
describe realistic mechanisms, including
slow charge noise and phonon-mediated inelastic
processes, which make this  parity-to-charge conversion 
scheme imperfect.
Our results provide guidelines for future experiments
aiming at demonstration and optimization of the
parity readout procedure.

The viewpoint we take is that quantum-information 
experiments with 1DTSs are expected in the near future, 
and therefore practical questions regarding feasible experiments
will arise.
For example, a relevant question is: based on the 
knowledge acquired in the previous decades
of coherent-control experiments with solid-state
qubit devices (superconducting qubits, spin qubits), 
what are the feasibility conditions for the upcoming 
topological quantum information experiments? 
Our work provides guidelines in that respect: see, e.g., 
the results summmarized in Fig.~\ref{fig:errorthresholds},
the map showing the proximity gap and
readout pulse strength required to achieve a certain
parity readout fidelity.

We note that in practice, a single 1DTS wire is not sufficient
for encoding a qubit, but at least two wires are required for 
that\cite{LeijnseReview}. 
Nevertheless, the minimal setup where parity-to-charge
conversion could be demonstrated is the single-wire setup
considered here. 
Extending our study to a two-wire scenario 
could reveal different error mechanisms, e.g., 
related to 
charge dipole formation\cite{Knapp_dephasing}.

The rest of the paper is organized as follows.
In section \ref{sec:paritytocharge}, we outline our model and
define the parity readout errror.
In section \ref{sec:errors}, we calculate the parity readout error
induced by three different error mechanisms:
leakage due to a strong readout tunnel pulse, 
incomplete charge Rabi oscillations due to slow charge noise,
and 
charge relaxation due to phonon emission. 
In a case study based on InAs heterostructure device parameters,
we estimate that the parity readout error is mainly due to
the first two error mechanisms. 
In section \ref{sec:discussion}, we discuss generalizations of
our results, and provide conclusions in section \ref{sec:conclusions}.

\section{Parity-to-charge conversion for a two-site Kitaev chain}
\label{sec:paritytocharge}

A simple model of a 1DTS,
often termed the \emph{Kitaev chain}, 
have been formulated by Kitaev\cite{KitaevChain}.
Besides serving as a minimal 1DTS model, it has been 
proposed that the the few-site Kitaev chain can be realized by
combining a few superconducting electrodes with a few
quantum dots\cite{LeijnsePoormans,SauChain,Fulga_2013}.

We start our analysis with the two-site Kitaev chain.
Together with a third site representing a quantum dot
used for charge readout (the \emph{readout dot} or simply \emph{dot},
see Fig.~\ref{fig:setup}b), 
our setup is described by the Hamiltonian
\begin{subequations}
\label{eq:twositekitaevchain}
\bean
H &=& H_\text{chain} + H_\text{dot} + H_\text{tun}
\\
H_\text{chain} &=& 
\varepsilon_1 c_1^\dag c_1 + 
\varepsilon_2 c_2^\dag c_2 \\
&+&  \nonumber
v(c_1^\dag c_{2} + h.c.)
+
\Delta(c_1^\dag c_2^\dag + h.c.)
\\
H_\text{dot} &=& \varepsilon_\text{dot} 
c^\dag_\text{dot} c_\text{dot} 
\\
H_\text{tun} &=& 
u(t) (c_1^\dag c_\text{dot} + c_2^\dag c_\text{dot} + h.c.)
\eean
\end{subequations}
Here, the Hamiltonian $H_\text{chain}$ of the Kitaev chain
is built up from the fermion creation an annihilation operators
($c$), 
and it includes the on-site energies $\varepsilon_1$ and
$\varepsilon_2$ of the chain, the normal hopping amplitude $v$, 
and the effective superconducting gap $\Delta$ $>0$.
The Hamiltonian $H_\text{dot}$ of the readout dot is parametrized by the 
on-site energy $\varepsilon_\text{dot}$.
Finally, the chain-dot tunneling Hamiltonian $H_\text{tun}$
is parametrized by the time-dependent hopping amplitude
$u(t)$, which is a non-negative real quantity throughout
this work.  
One possible generalization of our model is to take 
into account the spin degree of freedom.
\cite{PhysRevB.94.045316,PhysRevB.96.045440}

For now, we consider 
the \emph{ideal Kitaev limit},
where the tunneling amplitude within the chain matches
the effective superconducting gap,
$v=\Delta$, and the on-site energies
are zero, $\varepsilon_1 = \varepsilon_2 = 
\varepsilon_\text{dot} = 0$.
Later, we will describe the effect of disorder by
considering the on-site energies as random variables.
Following earlier work, we will consider parameter sets where the
effective superconducting gap is a few tens or hundreds of
microelectronvolts\cite{SauChain}, and the 
on-site disorder is of the order of $1 \, \mu$eV\cite{Camenzind}.

To describe the parity readout mechanism, we
use the complete Fock-space Hamiltonian here. 
However, often it is insightful and
sufficient to apply a low-energy 
description of the coupled chain-dot system\cite{FlensbergNonabelian}.
We summarize that alternative low-energy picture in 
Appendix \ref{app:lowenergy}.

Before specifying the parity readout procedure and
defining the parity readout error, 
we introduce a convenient basis
of the many-body states.
Since Eq.~\eqref{eq:twositekitaevchain} describes a 
spinless fermionic model with three sites, the dimension of the
full Fock space is 8. 
The Hamiltonian conserves the parity of fermion numbers, 
hence the dynamics is governed by two separate
$4\times 4$ Hamiltonians. 
It is convenient to use a product basis, 
composed of the energy eigenstates of $H_\text{chain}$
and $H_\text{dot}$. 
The energy eigenbasis of $H_\text{chain}$
is expressed in the occupation
number representation as 
\bean
\ket{\textrm{e}} &=& \frac{1}{\sqrt{2}}\left(\ket{11}-\ket{00}\right), \\
\ket{\textrm{e}'} &=& \frac{1}{\sqrt{2}}\left(\ket{11}+\ket{00}\right), \\
\ket{\textrm{o}} &=& \frac{1}{\sqrt{2}}\left(\ket{10}-\ket{01}\right), \\
\ket{\textrm{o}'} &=& \frac{1}{\sqrt{2}}\left(\ket{10}+\ket{01}\right),
\eean
whereas the eigenbasis of $H_\text{dot}$ 
will be denoted as $\ket{0}$ (empty dot) and 
$\ket{1}$ (filled dot).
Note that $\ket{\text{e}}$ and $\ket{\text{o}}$ are the 
even and odd ground states of $H_\text{chain}$, at energy
$-\Delta$, 
whereas $\ket{\text{e}'}$ and $\ket{\text{o}'}$ are the 
excited states, at energy $+\Delta$.

Using the product basis of
the even sector ordered as 
$\ket{\textrm{e},0},\; \ket{\textrm{o},1},\; \ket{\textrm{e}',0},\; \ket{\textrm{o}',1}$,
the  Hamiltonian matrix of this sector reads
\bean
\label{eq:hamiltonianeven}
H_\text{e} = \left(\bna{cccc}
- \Delta & u(t) & \frac 1 2 (\varepsilon_1 + \varepsilon_2) & 0 \\
u(t) & -\Delta + \varepsilon_\text{dot} & u(t) & \frac 1 2 (\varepsilon_1 - \varepsilon_2) \\
\frac 1 2 (\varepsilon_1 + \varepsilon_2) & u(t) & \Delta & 0 \\
0 & \frac 1 2 (\varepsilon_1 - \varepsilon_2) & 0 & \Delta +\varepsilon_\text{dot}
\eda \right).
\eean
Using the produce basis of the odd sector ordered 
as $\left\{\ket{\textrm{o},0},\; \ket{\textrm{e},1},\; \ket{\textrm{o}',0},\; 
\ket{\textrm{e}',1}\right\}$,
the Hamiltonian matrix of this sector reads
\bean
\label{eq:hamiltonianodd}
H_\text{o} = \left(\bna{cccc}
- \Delta & 0 & \frac 1 2 (\varepsilon_1 - \varepsilon_2) & 0 \\
0 & -\Delta + \varepsilon_\text{dot} & - u(t) & \frac 1 2 (\varepsilon_1 + \varepsilon_2) \\
\frac 1 2 (\varepsilon_1 - \varepsilon_2) & -u(t) & \Delta &u(t) \\
 0 & \frac 1 2 (\varepsilon_1 + \varepsilon_2) & u(t) & \Delta +\varepsilon_\text{dot}
\eda \right).
\eean
The eigenstates of the Hamiltonians
$H_\text{e}$ and $H_\text{o}$ for
$\varepsilon_1 = \varepsilon_2 = \varepsilon_\text{dot} =0$ and
$u(t) = 0$,
i.e., the basis states defined above,
are depicted in the energy level diagrams of 
Fig.~\ref{fig:levels}a and b, respectively. 
The matrix elements of the tunneling Hamiltonian $H_\text{tun}$ are also
depicted, as the orange arrows connecting the energy levels.

We envision the parity readout protocol as follows. 
Initally, the dot is empty, 
and we set the dot level in resonance with 
the chain chemical potential: $\varepsilon_\text{dot} = 0$.
The charge dynamics allowing for the readout is triggered by 
switching on the tunneling between each end of the
chain and the readout dot (Fig.~\ref{fig:setup}b), using the following 
squared-shape tunnel pulse (Fig.~\ref{fig:setup}d): 
\bean \label{eq:squarepulse}
u(t) = u_\textrm{max} \Theta(t)\Theta(\tau-t).
\eean
Here, $\tau$ is 
tunnel pulse duration, and $u_\textrm{max} > 0$ is the tunnel pulse strength.
We will also refer to $u_\textrm{max}$ as the readout speed.

Using the $4 \times 4$ matrix representations
of the Hamiltonians $H_\text{e}$ and $H_\text{o}$, 
it is straightforward to see how this pulse allows for the parity-to-charge
conversion. 
For the simple case when all on-site energies are zero,
the $2 \times 2$ low-energy block (top left block)
of $H_\text{e}$ has an off-diagonal element $u$, 
whereas the same block of $H_\text{o}$ has zero in the off-diagonal. 
That implies that if the system starts in the even ground state
$\ket{\text{e},0}$, then it will Rabi-oscillate to the state $\ket{\text{o},1}$,
but this Rabi oscillation does not happen when the initial state
is $\ket{\text{o},0}$. 
After a half Rabi oscillation, the parity of the wire is therefore
converted to the charge of the readout dot, which
can be measured by a proximal charge sensor \cite{Razmadzemajorana}.

\begin{figure*}
\centering
\includegraphics[width=2.0\columnwidth]{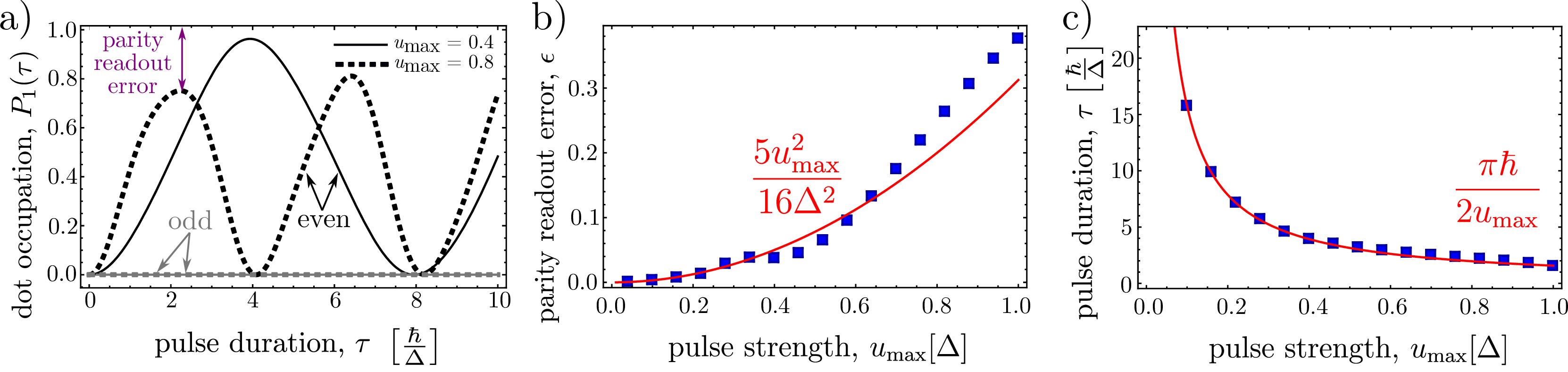}
\caption{
Parity readout error due to
leakage caused by a strong readout pulse.
(a) Time evolution of the occupation probability $P_1(\tau)$ of the 
readout dot, for the two different initial states
(even and odd), 
and for two different readout pulse strengths
(see legend).
(b) Parity readout error as a function of the readout pulse strength.
(c) Numerically obtained optimized readout pulse duration (boxes)
and its comparison to the analytical estimate (solid).
}
\label{fig:noiseless}
\end{figure*}

\begin{figure}
\includegraphics[width=1.0\columnwidth]{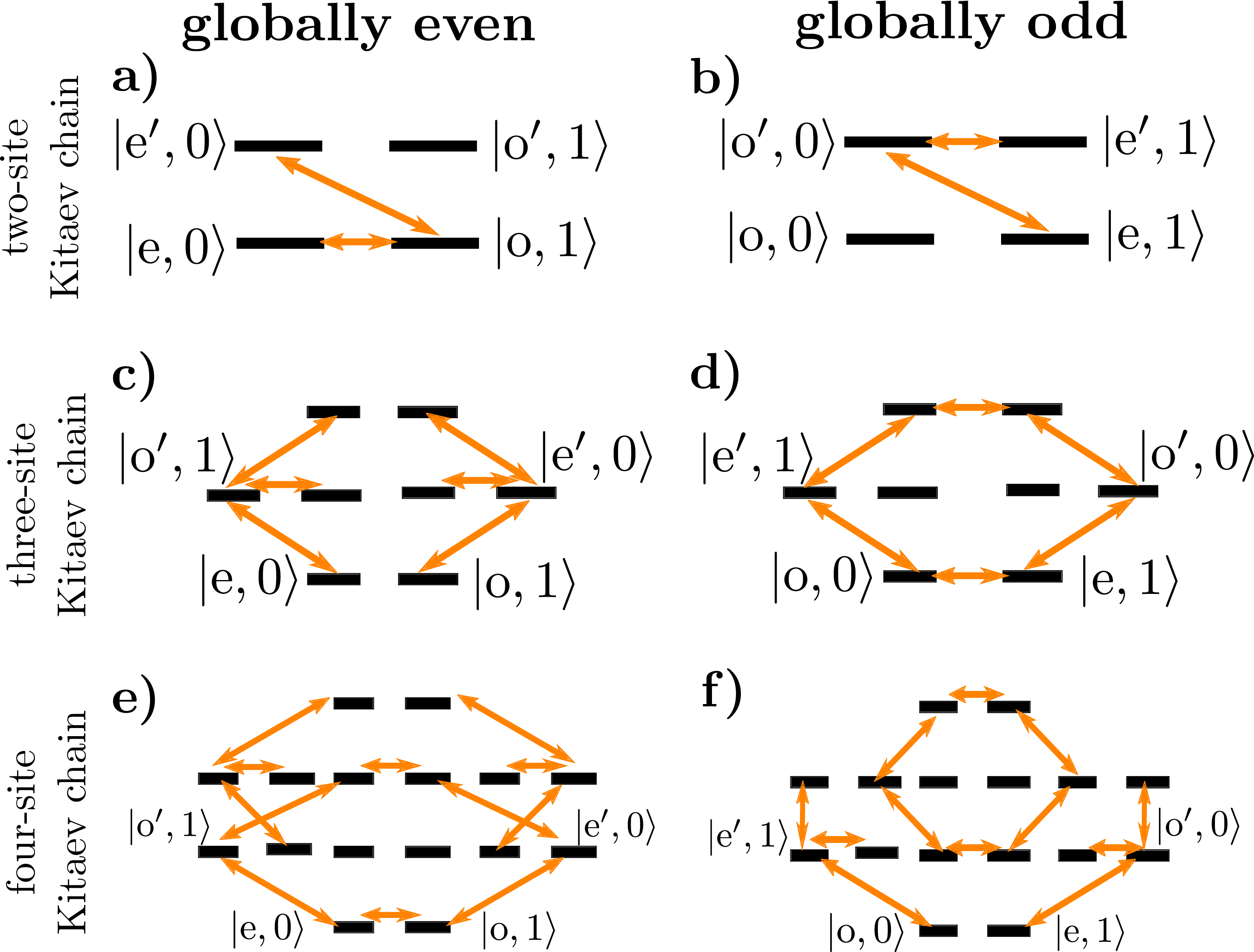}
\caption{
Energy levels and chain-dot tunneling matrix elements
in a simple model of parity readout.
Horizontal lines are the energy levels of
the composite system of the Kitaev chain and 
the readout dot, described by $H_\text{chain} + H_\text{dot}$. 
Orange arrows are the nonzero tunneling matrix elements
of $H_\text{tun}$.
}
\label{fig:levels}
\end{figure}

To evaluate the quality 
of the parity readout, we introduce the 
\emph{parity readout error} $\epsilon$ as follows. 
We assume that the initial state is the one where
the chain is in one of its ground states, 
either in the even or in the odd state, 
and the quantum dot is empty.
The readout procedure consists of a tunneling
time window and the subsequent charge measurement.
Using our model, we can calculate how the probabilities
of finding the quantum dot empty or filled upon charge
measurement depends on the initial state, denoted as
$P_{0 \leftarrow \textrm{e}}$,
$P_{1 \leftarrow \textrm{e}}$, 
$P_{0 \leftarrow \textrm{o}}$,
$P_{1 \leftarrow \textrm{o}}$.
These probabilities fulfill 
the normalization conditions
$P_{0 \leftarrow \textrm{e}} + P_{1 \leftarrow \textrm{e}} = 1$
and 
$P_{0 \leftarrow \textrm{o}} + P_{1 \leftarrow \textrm{o}} = 1$, 
therefore 
only 2 out of these 4 quantities are independent.
Ideally, the charge measurement gives a clear distinction
between the two different initial states, i.e., 
$P_{0 \leftarrow \textrm{e}} = P_{1 \leftarrow \textrm{o}} = 0$
and
$P_{0 \leftarrow \textrm{o}} = P_{1 \leftarrow \textrm{e}} = 1$.
The non-ideal nature of the readout procedure is 
therefore characterized by 
the \emph{parity readout error}
\bean
\label{eq:epsilon}
\epsilon = \max\{
P_{0 \leftarrow \textrm{e}},P_{1 \leftarrow \textrm{o}}
\}.
\eean

The concept of the parity readout error
is exemplified using Fig.~\ref{fig:noiseless}a.
Focus on the time evolution shown by the
dashed lines. 
The black (gray) dashed line shows the dot 
occupation probability as the function 
of the readout pulse duration for the 
even (odd) initial state, 
for a two-site Kitaev chain in the ideal limit.
These curves reveal that $P_{1\leftarrow \textrm{o}} \approx 0$ 
whereas
$P_{1 \leftarrow  \textrm{e}} \approx 0.75$.
The latter implies  
$P_{0 \leftarrow  \textrm{e}} \approx 0.25 \gg P_{1\leftarrow \textrm{o}}$, 
hence the parity readout error defined in
Eq.~\eqref{eq:epsilon} is
$\epsilon = P_{0 \leftarrow \textrm{e}}$.
E.g., a pulse duration $\tau \approx 2 \hbar / \Delta$
implies a parity readout error $\epsilon \approx 0.25$.
In the rest of the paper, 
we will use $\epsilon$ to quantify the imperfect
nature of the readout procedure. 

To be more precise, the parameter $\epsilon$ 
characterizes \emph{single-shot} readout error.
This is the relevant figure of merit for many
quantum-information protocols, including 
quantum error correction\cite{NielsenChuang},
sampling algorithms\cite{Arute}, 
and measurement-based quantum computing, e.g., one-way quantum computing. \cite{Raussendorf}

There are a few different proposals for the readout of Majorana qubits, 
and some of them were suggested to 
enjoy exponential protection against certain error 
mechanisms, e.g., the one based on continuous charge measurement
of an auxiliary dot\cite{KarzigScalable,Plugge_2017,Steiner,MunkReadout}.
The Rabi-oscillation-based scheme\cite{Gharavi,Plugge_2017} we study 
here
requires fine tuning of pulse shape and timing, 
and hence hence does not enjoy exponential 
protection against, e.g., such pulse errors. 
Nevertheless, there are strong reasons motivating the 
analyis of such non-protected protocols, including the
one presented in this work.

(1) \emph{Experiments}. 
There are many experiments probing charge Rabi oscillations
in quantum-dot systems, see, e.g., 
Refs.~\onlinecite{Hayashi,Petersson,ZhanShi};
this is a well established experimental setting in the field of quantum
devices. 
The readout scheme studied here is a natural extension, 
and our results can help the design and the analysis of such a future
experiment.

(2) \emph{A strong measurement is a simple measurement.}
One advantage of the Rabi-oscillation-based readout scheme
studied here is its conceptual simplicity, as compared to 
exponentially protected schemes, e.g., based on continuous
charge readout\cite{KarzigScalable,Plugge_2017,Steiner,MunkReadout}.
Our scheme offers a straightforward way to calculate 
the probabilities of the measurement outcomes, even
in the presence of quasiparticle excitation 
(aka \emph{leakage}, see section \ref{sec:leakage}), as it uses
a strong, von Neumann-type measurement in a setting where
the readout dot is decoupled from the Kitaev chain. 
On the other hand, in the exponentially protected
readout scheme (e.g., continuous charge readout),
the analysis must incorporate a measurement theory, that is, 
a detailed description of the interaction of the measurement
device with the measured system,
(e.g., electrons flowing in a quantum point 
contact whose current reveals the charge of the readout dot).
Recent theory works\cite{Steiner,MunkReadout} make important
steps in that direction, but - perhaps in part due to the complexity
of the measurement model - those works do not incorporate
charge noise, do not incorporate quasiparticle excitation errors 
(leakage), and do not provide results for the readout fidelity.

(3) \emph{Rapid measurement.}
In an experiment, when the system size and the excitation gap are
both fixed and finite, decoherence mechanisms (e.g., quasiparticle
poisoning) impose an upper bound on the readout time.
The protocol we study here is fast, 
enabling readout before decoherence happens. 
This can be an advantage with respect to adiabatic protocols, which, in 
principle, enjoy exponential protection in the adiabatic limit, 
but decoherence might obstruct reaching that limit.

\section{Error mechanisms for a two-site Kitaev chain}
\label{sec:errors}

In this section, we quantify the imperfect nature of the
parity-to-charge conversion, using the simplest
possible model for the 1DTS, the two-site Kitaev chain
introduced in Eq.~\eqref{eq:twositekitaevchain}. 
In the three subsections, we describe three mechanisms:
(A) leakage, (B) incomplete charge Rabi oscillations,
and (C) charge relaxation. 
These arise due to 
(A) a strong tunnelling pulse, 
(B) slow charge noise inducing a random on-site
energy component on the readout dot, and
(C) phonon-mediated inelastic processes.

Our motivation for a detailed study of the two-site Kitaev chain
as a minimal model is threefold. 
First, this is a minimal model where the derivations can be 
followed easily. 
Second, it turns out that this minimal model exhibits and exemplifies
all error mechanisms we see on longer chains, and
the order-of-magnitude of errors does not depend
significantly on the chain length, see Figs. 5, 7 and 8 below.
Third, an analysis of the two-site Kitaev chain model
is important for future experiments where a Kitaev chain 
is built from quantum dots\cite{LeijnsePoormans,SauChain,Fulga_2013}.
There, the simplest device is of course a two-site chain;
it is not expected to show topological physics, but is a natural
and important building block to study.

\subsection{Leakage due to strong readout tunnel pulse}
\label{sec:leakage}

Even in the absence of noise, the tunnel pulse 
between the 1DTS and the readout dot 
induces transitions between the low-energy states
and high-energy states having finite-energy quasiparticles
in the chain.  
This \emph{leakage} gets stronger for stronger
tunnelling pulse. 

First, we illustrate this leakage mechanism and its dependence
on the tunnel pulse strength $u_\textrm{max}$ in Fig.~\ref{fig:noiseless}a.
The gray curves in Fig.~\ref{fig:noiseless}a
show the readout dot occupation after a tunneling pulse
of duration $\tau$, for the case when the chain 
is in its odd state initially. 
For both the weaker (solid) and stronger (dashed)
tunnel pulse, $P_1(\tau) = 0$, which 
is perfect for the parity-to-charge conversion. 
The black curves show the readout dot occupation 
after a tunneling pulse of duration $\tau$, for the case
when the chain is in its even state initially. 
Naturally, 
the stronger pulse (dashed) provides a faster readout, 
with an optimal pulse duration of $\tau \approx 2 \hbar / (0.8 \Delta)$, 
compared to the weaker pulse (solid) with 
an optimal pulse duration of $\tau \approx 2 \hbar / (0.4 \Delta)$. 

Another feature of the black curves is that there is a
finite parity readout error for both tunneling pulse strengths,
the stronger pulse resulting in a greater error
(marked with the `parity readout error' label).
The reason for this error is leakage, that is, 
that the high-energy excited
states of the system acquire a finite population during
the tunnel-induced dynamics, and they get
more populated if we choose a stronger tunnel pulse. 
This role of the excited states is illustrated in 
Fig.~\ref{fig:levels}a, where the orange arrows depict
the tunneling-induced matrix elements between
the eigenstates of the uncoupled wire-dot Hamiltonian.
This level diagram shows that the state
$\ket{\textrm{o},1}$ is tunnel-coupled to the high-energy excited
state $\ket{\textrm{e}', 0}$, leading to increasing leakage with
increasing tunnel pulse strength.

To quantify the parity readout error due to leakage, 
we determine its optimum for various values 
of the tunnel pulse strength. 
We numerically simulate the time evolution of
the parity-to-charge conversion process to obtain
the readout dot occupation $P_{1\leftarrow \textrm{e}}(\tau)$, 
and numerically find its first maximum, providing
optimal parity readout. 
This minimal error is shown as the 
discrete data set (boxes) in 
Fig.~\ref{fig:noiseless}b as a function of the
tunnel pulse strength $u_\textrm{max}$, revealing a clear 
trend that a slower readout is more
reliable. 

A perturbative analytical treatment of the dynamics
(see appendix \ref{app:leakage}) valid for
$u_\textrm{max} \ll \Delta$ yields an analytical result for
the parity readout error,
\bean
\label{eq:epsilontwosite}
\epsilon = \frac{5}{16} \left(\frac{u_\textrm{max}}{\Delta}\right)^2.
\eean
This function is plotted as the solid line in 
Fig.~\ref{fig:noiseless}, showing a reasonable agreement
with the numerical data set, especially in the 
slow readout range, $u_\textrm{max} \ll \Delta$, where the perturbative 
treatment is justified.
Note that the numerically optimized tunnel pulse
durations, shown as the data set (boxes)
in Fig.~\ref{fig:noiseless}c, show hardly any deviations
from the analytical expression 
$\tau_\text{ideal} = \pi \hbar / (2 u_\text{max})$
(solid line) derived for a  
two-level Rabi oscillation.

\subsection{Incomplete charge Rabi oscillations due to slow charge noise}
\label{sec:slowchargenoise}

A generic feature of mesoscopic electronic devices 
is the presence of electric potential fluctuations\cite{Camenzind}, 
often characterized by
by a frequency-dependent power spectrum $S(f)$ 
proportional to $1/f$ \cite{Freeman, Petit,Yoneda2018}.

\begin{figure}
\includegraphics[width=1.0\columnwidth]{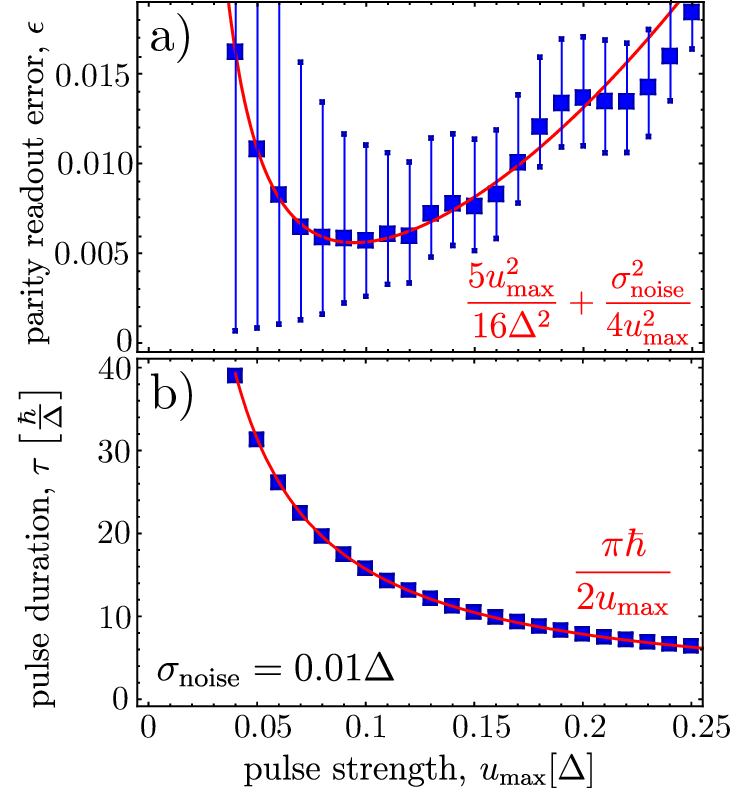}
\caption{
Combined effects of tunnel pulse strength and slow charge noise
on parity readout error.
(a) Parity readout error averaged numerically for 
5000 disorder realizations (boxes) and calculated analytically
using perturbation theory (solid). 
(b) Optimal readout pulse duration 
obtained from the numerical optimization (boxes) and from
the analytical estimate
(solid).}
\label{fig:noisy}
\end{figure}

Due to the dominance of the low-frequency component, 
we will refer to this type of noise as \emph{slow charge noise},
and, following earlier works \cite{XinWang,Tosi2017,Watson,Boter}, will  use a quasistatic
model to describe it. 
In our model, we account for the most prominent effect of this
noise, which is to detune the energy level of the readout dot.
Because of this detuning, charge Rabi oscillations are only 
partial, and hence the parity-to-charge conversion is
imperfect. 
As we show (Fig.~\ref{fig:noisy}a),
the main effect of the slow charge noise is 
that the parity readout error increases as the 
tunnel pulse strength is decreased.
Given the fact that the leakage error increases 
with increasing tunnel pulse strength, our conclusion 
is that in the presence of slow charge noise, there is
an \emph{optimal} tunnel pulse strength 
that minimizes the parity readout error. 

To reach that conclusion, we assume that the
on-site energies of both the Kitaev chain and the readout
dot are random.
Furthermore, we model slow charge noise
as quasistatic, by which we mean that
its configuration is time-independent for any single
run of the experiment, and changes randomly 
between different runs.
For example, in the case of a two-site Kitaev chain, 
we model the three
on-site energies $\varepsilon_1$, $\varepsilon_2$ and
$\varepsilon_\text{dot}$
as independent random Gaussian variables with 
zero mean and the same standard deviation 
$\sigma_\textrm{noise}$. 

We mimic experimental conditions by first 
optimizing the parity readout scheme for the tunnel 
pulse duration, and assuming
that the system is subject to noise already during
this optimization. 
First, we generate $N_\text{r} = 5000$
random realizations of the on-site energy vector
$\xi = (\varepsilon_1,\varepsilon_2,\varepsilon_\text{dot})$,
which we call disorder realizations from now on,
and denote as $\xi_j$ ($j \in \{1,2,\dots, N_\text{r}\})$.
Second, we simulate the coherent dynamics of
the parity readout procedure, i.e., the charge
Rabi oscillation, for each disorder realization,
yielding the disorder-dependent occupation
probabilities $P_{1\leftarrow \textrm{e}}(t,\xi)$
and $P_{1\leftarrow \textrm{o}}(t,\xi)$ of the readout dot.
Third, we evaluate the disorder-averaged
readout dot occupation probabilities, 
$\bar{P}_{1\leftarrow \textrm{e}}(t) = (1/N_r) \sum_{j = 1}^{N_\text{r}}
P_{1\leftarrow \textrm{e}}(t;\xi_j)$ and
$\bar{P}_{1\leftarrow \textrm{o}}(t) = 
(1/N_r) \sum_{j = 1}^{N_\text{r}}
P_{1\leftarrow \textrm{o}}(t;\xi_j)$.
Finally, we numerically find the shortest time $t$ where
the parity readout error has a local minimum as the function
of $t$; this we call the optimized tunnel 
pulse duration $\tau_\text{opt}$.

The optimized tunnel pulse duration 
$\tau_\text{opt}$
is of course a function of the tunnel pulse strength
$u_\textrm{max}$. 
This functional dependence $\tau_\text{opt}(u_\textrm{max})$ 
is shown in Fig.~\ref{fig:noisy}b as the data points (blue boxes),
for the case when the noise strength is
$\sigma_\textrm{noise} = 0.01 \Delta $,
computed from $N_\text{r} = 5000$ realizations.
For comparison, we also show the function
$\tau_\text{ideal}(u_\textrm{max}) = \frac{\pi \hbar}{2u_\textrm{max}}$, 
which describes the optimal
tunnel pulse duration expected from idealized, complete
two-level charge Rabi oscillations.
We conclude that there is no observable difference between
the numerically optimized
$\tau_\text{opt}$ and analytically estimated $\tau_\text{ideal}$
values of the optimal tunnel pulse duration. 

Next, we simulate the effect of slow charge noise on the
parity readout error. 
Since slow charge noise is modelled as a
random vector $\xi$ of the on-site potentials, 
the parity readout error is also a random variable.
We consider a specific value for the noise strength 
$\sigma_\textrm{noise} = 0.01 \Delta$, fix the tunnel pulse strength 
$u_\textrm{max}$, and apply a tunnel pulse of duration
$\tau_\text{opt}(u_\textrm{max})$.
Then, the parity readout error is characterized by a 
probability density function $\rho(\epsilon;u_\textrm{max})$.
The data points (blue boxes) in Fig.~\ref{fig:noisy}a 
show the average parity readout error as a function 
of the tunnel pulse strength $u_\textrm{max}$, as obtained from 
finite-sample averaging with $N_\text{r} = 5000 $ 
disorder realizations. 
The error bars in Fig.~\ref{fig:noisy}a connect
the 1st and 9th deciles of the $N_\text{r}$ 
parity readout values.

The average parity readout error shows a non-monotonic
behavior as $u_\textrm{max}$ is increased.
For slow readout, that is, small $u_\textrm{max}$, 
the parity readout error grows as the readout
speed is reduced. 
For fast readout, the parity readout error grows as the
readout speed is increased. 
The latter feature is caused by leakage, as discussed
in the previous subsection. 
The former feature is a consequence of the disorder.
More precisely, in the weakly
disordered regime ($\sigma_\textrm{noise} \ll \Delta $)
studied here, 
the only relevant component of the disorder is the
random on-site energy of the readout dot.
The randomness of the on-site energies of the Kitaev chain
has hardly any effect on the results, since the energy detuning suffered
by the chain eigenstates are second order in the disorder, 
as seen by the fact that $\varepsilon_1$ and $\varepsilon_2$
does not appear in the diagonals of the matrices
$H_\text{e}$ and $H_\text{o}$ of Eqs.~\eqref{eq:hamiltonianeven}
and \eqref{eq:hamiltonianodd}.

In fact, the simple structure of the Hamiltonians $H_\text{e}$ and
$H_\text{o}$ 
allows us to derive an analytical estimate for
the parity readout error, in the limit
when $\sigma_\textrm{noise} \ll u_\textrm{max} \ll \Delta$. 
The derivation, based on textbook perturbation theory, 
is outlined in appendix \ref{app:slowchargenoise}.
The resulting formula for the parity readout error
is 
\bean
\label{eq:twositeanalytical}
\epsilon = \frac{5}{16} \left(\frac{u_\text{max}}{\Delta}\right)^2
+ \frac{1}{4} \left(\frac{\sigma_\textrm{noise}}{u_\text{max}}\right)^2
\eean
This analytical result is shown as the solid line in 
Fig.~\ref{fig:noisy}a, providing a reasonable
approximation of the numerical data (blue boxes).
Note that the perturbative calculation in appendix \ref{app:slowchargenoise}
reveals that the parity readout error in this two-site model
is dominated by the first-order effect that the on-site energy
of the readout dot is detuned by charge noise. 
Majorana hybridization is only a second-order effect\cite{LeijnsePoormans},
i.e. a minor correction of the error caused by the 
readout dot detuning.

One important practical conclusion drawn from 
the results of Fig.~\ref{fig:noisy}a is that
the readout speed $u_\text{max}$ should be fine tuned to 
a finite value (around $0.1 \Delta$ in this example) to 
minimize the readout error. 
In fact, the optimal readout speed can be found by
using Eq.~\eqref{eq:twositeanalytical} and 
solving $d \epsilon / d u_\text{max} = 0$, yielding
\bean
\label{eq:optimalspeed2}
u_\text{max}^{\text{(opt)}} = \frac{\sqrt{2}}{5^{1/4}} \sqrt{\sigma_\textrm{noise} \Delta},
\eean
and the corresponding optimized (minimized) parity readout error
is 
\bean
\label{eq:optimalerror2}
\epsilon^\text{(opt)} = \frac{\sqrt{5}}{4} \frac {\sigma_\textrm{noise}}{ \Delta}.
\eean

\begin{figure}
\centering
\includegraphics[width=1.0\columnwidth]{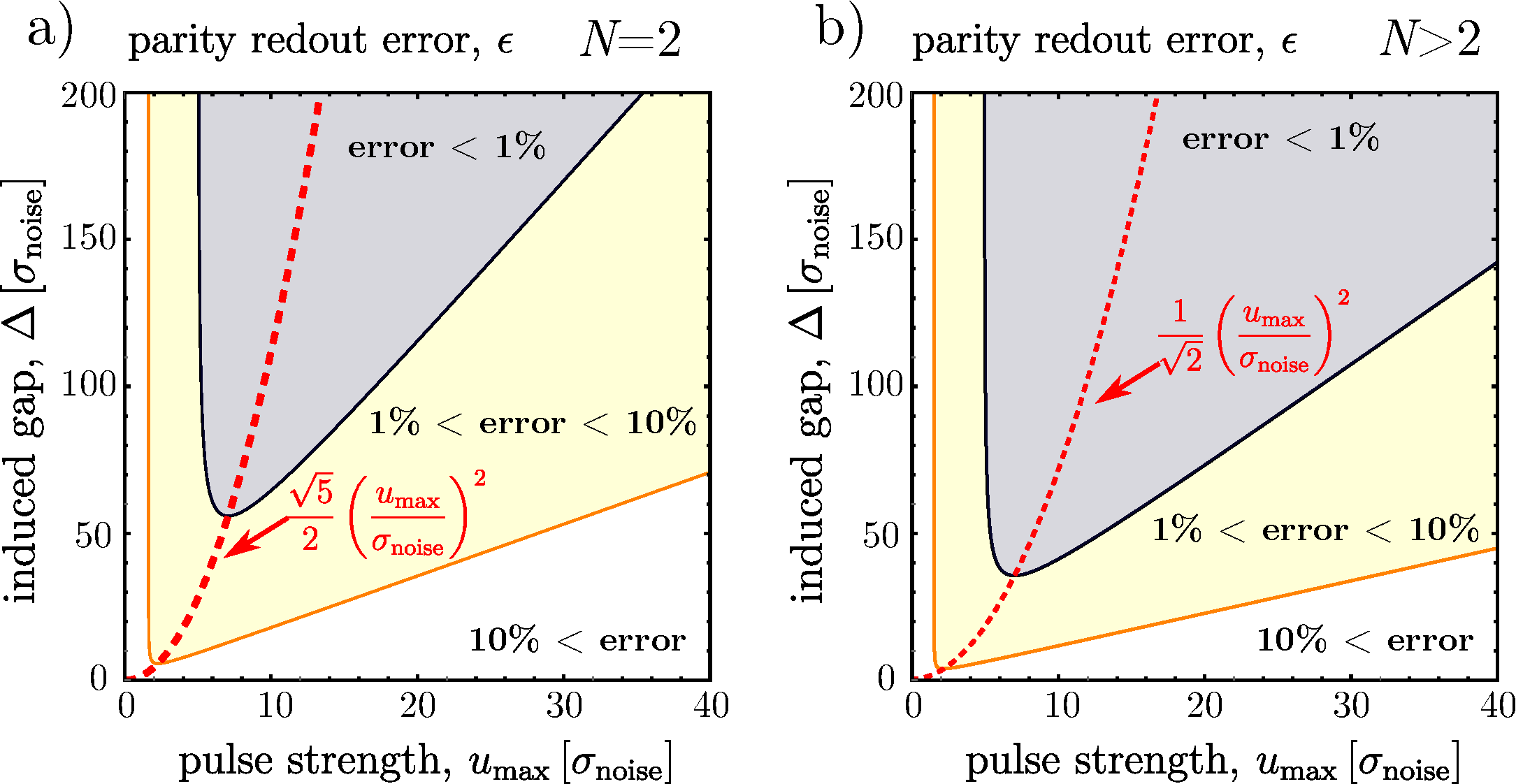}
\caption{
Requirements on the system parameters to achieve
certain parity readout error thresholds.
Evaluated from the perturbative analytical result
Eq.~\eqref{eq:twositeanalytical}. The dashed red lines show the optimal pulse strength, derived by inserting the relations in 
Eqs.~\eqref{eq:optimalspeed2} and 
\eqref{eq:optimalspeedlong}, respectively.
}
\label{fig:errorthresholds}
\end{figure}

A practical implication of our result \eqref{eq:twositeanalytical} 
is that it provides a map 
of realistic parameter values that are required for a reasonably
accurate parity readout experiment. 
This is shown in Fig.~\ref{fig:errorthresholds}a, where the
different regions correspond to different ranges of 
parity readout errors.
For example, 
taking the realistic noise strength value\cite{Freeman,Camenzind} $\sigma_\textrm{noise} = 1\, \mu\text{eV}$,
the figure indicates that an effective superconducting gap  
$ \Delta \gtrsim 55 \, \mu\text{eV}$ 
and a tunnel pulse strengh $\sim 10\, \mu$eV
(corresponding to tunnel pulse duration $\sim 100$ ps)
are required to have a parity readout error below 1 percent.
This sounds in principle achievable, e.g., with the commonly used
Al-coated III-V semiconductor nanowires \cite{Krogstrup2015,Lutchyn2018}
where the Al gap is a few hundred microelectronvolts. 
Finally, the dashed line in Fig.~\ref{fig:errorthresholds}a
shows the optimal readout 
tunnel strength as a function of the induced gap $\Delta$.

\subsection{Charge relaxation due to phonon emission}
\label{sec:phonon}

In this section, we describe how  parity readout
error is induced due to spontaneous phonon emission,
even at zero temperature. 
We use a simple example to illustrate the role of phonons:
a three-dimensional heterostructure based on InAs
as the semiconducting material\cite{Shabani,Suominen,Nichele}.
Because of the available phonon states, 
the Rabi oscillation will not be fully coherent, but 
a charge relaxation process assisted by the spontaneous emission
of a phonon will drive the system toward a quantum dot
state that is a mixture of empty and filled, and hence
will corrupt the parity-to-charge conversion process.  
We calculate the phonon-induced error due to the 
deformation-potential electron-phonon coupling
in this particular setup, finding 
that the error increases with increasing tunnel 
pulse strength as $\epsilon \propto u_\textrm{max}^2$,
similarly to the leakage error discussed in 
section \ref{sec:leakage}, but we estimate that the phonon-induced
error is much smaller than the leakage-induced error. 
We also 
note that the results of the phonon analysis depend qualitatively on
the sample dimensionality and the electron-phonon interaction mechanism,
we study only one example case here.

\begin{figure}
\includegraphics[width=1.0\columnwidth]{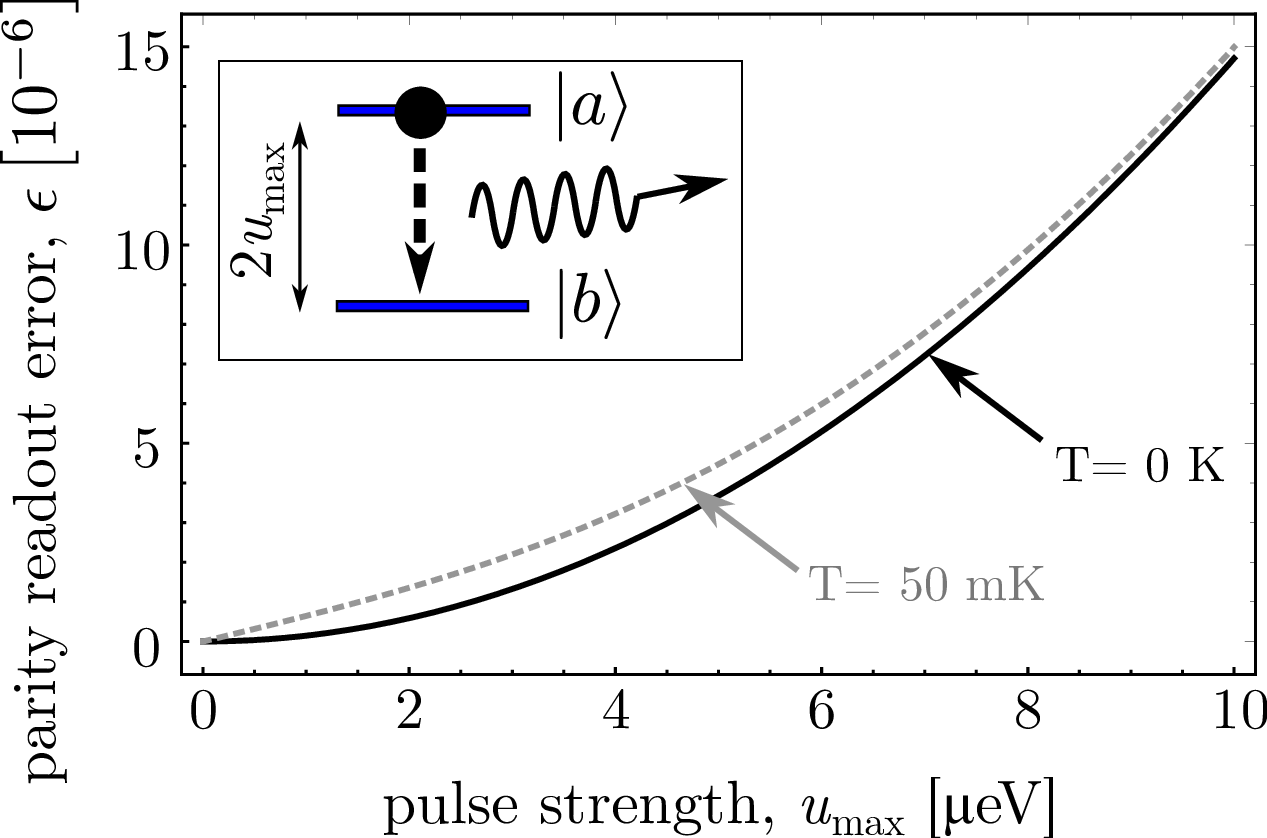}
\caption{
Parity readout error 
due to phonon-mediated 
charge relaxation.
At $T = 0$ K, the error is caused by spontaneous phonon emission
(inset). 
At finite $T$, spontaneous and stimulated emission as well as
absorption of thermal phonons contribute to the error.
}
\label{fig:phonon}
\end{figure}

To illustrate the origin of the phonon-induced error, 
consider the case of zero temperature. 
First, take the case when the wire parity
is even, corresponding to the level structure shown in Fig.~\ref{fig:levels}a.
In this case, when the tunnel pulse is turned on,
the instantaneous energy eigenstates are 
the bonding 
$\ket{b} = \frac{1}{\sqrt{2}} \left(
\ket{\textrm{e},0} - \ket{\textrm{o},1}
\right)$
and antibonding
$\ket{a} = \frac{1}{\sqrt{2}} \left(
\ket{\textrm{e},0} + \ket{\textrm{o},1}
\right)$
combinations of the 
basis states, which are separated by an energy gap of 
$2 u_\textrm{max}$, as shown in the inset of Fig.~\ref{fig:phonon}.
The initial state $\ket{\textrm{e},0}$ is a balanced superposition of the
bonding and antibonding states, that is, with a probability of 1/2
it occupies the excited state $\ket{a}$, and hence can
relax to $\ket{b}$ by emitting a phonon.
This relaxation process ends with the state $\ket{b}$
that has only $e/2$ charge on the readout dot, instead of charge
$e$ in the ideal case, implying that
this phonon emission induces an error in the parity-to-charge
conversion.

We describe the error due to this relaxation mechanism in two
steps.
First, we use the Bloch-Redfield master equation\cite{Makhlin}
to relate the energetically downhill $\ket{a} \mapsto \ket{b}$
relaxation rate $\Gamma_\downarrow$
to the parity readout error $\epsilon$.
Second, we estimate the relaxation rate $\Gamma_\downarrow$
induced by the deformation-potential electron-phonon
interaction mechanism in an InAs heterostructure, 
using Fermi's Golden Rule, in a fashion similar 
to spin relaxation calculations in quantum dots\cite{Khaetskii}. 

To describe parity readout error due to 
relaxation, we focus on the two-dimensional
subspace spanned by $\ket{b}$ and $\ket{a}$.
The density matrix describing the
initial state is $\rho(t=0) = \ket{\textrm{e},0}\bra{\textrm{e},0}$. 
Parity readout error is the probability of finding the readout dot
empty after the tunnel pulse duration $\tau = \pi \hbar / (2 u_\textrm{max})$,
that is, 
\bean
\label{eq:phononerror}
\epsilon = \text{Tr} \left[
\ket{\textrm{e},0}\bra{\textrm{e},0} \rho(\tau)
\right].
\eean
The time evolution of the density matrix 
is governed by the zero-temperature Bloch-Redfield equations:
\begin{subequations}
\bean
\dot{\rho}_{aa}(t) &=& - \Gamma_\downarrow \rho_{aa}(t), \\
\dot{\rho}_{bb}(t) &=& \Gamma_\downarrow \rho_{aa}(t), \\
\dot{\rho}_{ba}(t) &=& \frac{2i u_\textrm{max}}{\hbar} \rho_{ba}(t) 
- \frac 1 2 \Gamma_\downarrow \rho_{ba}(t).
\eean
\end{subequations}
Solving this and inserting the solution to Eq.~\eqref{eq:phononerror}
yields 
\bean
\label{eq:phononerror2}
\epsilon = 
\frac 1 2 \left(1 - e^{- \pi \hbar \Gamma_\downarrow/(4 u_\textrm{max})}\right)
\approx
\frac{\hbar \pi}{8 u_\textrm{max}} \Gamma_\downarrow,
\eean
where we assumed that the exponent is much smaller than one, 
to be confirmed below. 

Now, we determine the dependence of the $\ket{a} \mapsto \ket{b}$
relaxation rate
$\Gamma_\downarrow$ on the tunnel pulse strength $u_\textrm{max}$. 
Since we envision that the inverse time scale
of the readout (i.e., the charge Rabi frequency) 
is well below the THz range of typical
optical-phonon frequencies, it is safe to assume
that the phonon emitted upon the relaxation process is
an acoustic one. 
We will use InAs as the exemplary host material, 
and for simplicity, we assume that the interaction between the
electrons and the phonons
is described by the deformation-potential mechanism, 
and that the acoustic phonons can be characterized as
longitudinal and transverse bulk phonons.
Note that under these assumptions, the transverse
acoustic phonons do not cause volume change, and
therefore they do not contribute to the deformation potential. 

The electron-phonon Hamiltonian reads\cite{YuCardona}
\bean
H_\text{e-ph} = \Xi \text{Tr} \varepsilon(\vec r),
\eean
where $\Xi$ is the deformation-potential constant, 
and $\varepsilon(\vec r)$ is the strain tensor 
at the position $\vec r$ of the electron.
Using a single-electron basis formed by spatially localized states
on the three sites ($1$, $2$, and `dot') of our model 
[Eq.~\eqref{eq:twositekitaevchain}], the e-ph contribution 
to the globally even $4\times 4$ sector of the $8\times 8$
Hamiltonian reads:
\bean
H_\text{e-ph} = 
\left(
\bna{cccc}
0 & 0 & \frac{\tilde \varepsilon_1 + \tilde \varepsilon_2}{2} & 0 \\
0 & \tilde \varepsilon_\text{dot} & 0 & \frac{\tilde \varepsilon_1 
- \tilde\varepsilon_2}{2} \\
\frac{\tilde \varepsilon_1 + \tilde \varepsilon_2}{2} & 0 & 0 & 0 \\
0 & \frac{\tilde \varepsilon_1 - \tilde \varepsilon_2}{2} & 0 & \tilde \varepsilon_\text{dot}
\eda \right)
\label{eq:eph4x4}
\eean
where we have 
used the basis $\ket{\textrm{e},0}$, $\ket{\textrm{o},1}$, 
$\ket{\textrm{e}', 0}$, $\ket{\textrm{o}',1}$.
Furthermore, we assume that the deformation can be approximated
as homogeneous within each site of the model, and hence 
the phonon-induced on-site energy shifts read, e.g., as
$\tilde \varepsilon_1 \equiv \Xi \, \text{Tr}[\varepsilon(\vec r_1)]$.

As seen in Eq.~\eqref{eq:eph4x4}, the phonon-induced on-site
energy shifts on the two-site Kitaev
chain do not give direct matrix elements within the 
relevant subspace spanned by $\ket{\textrm{e},0}$ and $\ket{\textrm{o},1}$, 
hence it is reasonable to neglect those and keep only 
the more important terms proportional to $\tilde \varepsilon_\text{dot}$.
That is, for our purposes, the Hamiltonian is approximated 
as
\bean
H_\text{e-ph} \approx \tilde \varepsilon_\text{dot} 
\ket{\textrm{o},1} \bra{\textrm{o},1}
\equiv \frac{\tilde \varepsilon_\text{dot} }{2}
\left(
\ket{a}\bra{b} + h.c.
\right).
\eean
From this form, it is clear that the e-ph interaction 
induces a transition from $\ket{a}$ to $\ket{b}$.
We calculate the rate of this transition from Fermi's golden rule:
\bean
\label{eq:fgr}
\Gamma_\downarrow =
\frac{2\pi}{\hbar} \sum_{\vec q_f}
\left|
\braket{b,\vec q_f |  H_\text{e-ph} | a,0}
\right|^2 \delta(2u_\textrm{max} - \hbar c_L q_f),
\eean
where $0$ stands for the phonon vacuum, $\vec q_f$ is
the wave vector of the phonon emitted upon the transition, 
$c_L$ is the speed of sound for longitudinal phonons, 
and we applied a linear, isotropic approximation for the
dispersion of the acoustic phonons, $\omega_{\vec q} \approx c_L q$. 

To evaluate the matrix element in Eq.~\eqref{eq:fgr},
we recall that the e-ph Hamiltonian includes
$
\tilde \varepsilon_\text{dot} = 
\Xi \text{Tr} [\varepsilon(\vec r_\text{dot})]$, 
where the strain tensor is related to the displacement field $\vec u(\vec r)$
via $\varepsilon_{jk} = (\partial_j u_k + \partial_k u_j)/2$,
and the displacement field of the longitudinal
acoustic phonons is represented by the phonon creation and
annihilation 
operators via
\bean
\vec u(\vec r) = \sum_{\vec q} \sqrt{\frac{\hbar}{ 2 \rho V \omega_{\vec q}}}
\left(a_{\vec q} + a^\dag_{-\vec q} \right) 
e^{i \vec q \vec r}  \vec q /q.
\eean
Here, $\rho$ is the mass density of the host material and 
$V$ is the volume of the host crystal.
This expression allows a straightforward evaluation of Eq.~\eqref{eq:fgr}
in the $V\to \infty$ limit, yielding
\bean
\label{eq:gammadown}
\Gamma_\downarrow = \frac 1 \pi 
\frac{\Xi^2 u_\textrm{max}^3}{\rho \hbar^4 c_L^5}.
\eean

Using the result \eqref{eq:gammadown}
in the error formula Eq.~\eqref{eq:phononerror2}, we obtain
\bean
\label{eq:phononerror3}
\epsilon = \frac{\Xi^2 u_\textrm{max}^2}{8\rho \hbar^3 c_L^5}.
\eean
The error depends quadratically on the tunnel pulse strength
$u_\textrm{max}$, in a fashion similar to the leakage error evaluated in
Eq.~\eqref{eq:epsilontwosite}.
In our numerical example however, the phonon-induced error is
much smaller than the leakage error: 
for $\Delta = 100 \, \mu$eV and $u_\textrm{max} = 20\, \mu$eV, 
the leakage error is $\epsilon \approx 1.25 \times 10^{-2}$, 
whereas for the InAs parameters 
$\rho = 5667\, \text{kg}/\text{m}^3$,
$c_L = 4430\, \text{m}/\text{s}$, and
$\Xi = 4.5$ eV,
the phonon-induced error is
$\epsilon \approx 6 \times 10^{-5}$.
More generally, 
the form of Eq.~\eqref{eq:phononerror3} corresponding 
to these InAs parameters reads
\bean
\epsilon = \left(\frac{u_\textrm{max}}{2.61\, \text{meV}}\right)^2.
\eean
This result is shown as the solid line in Fig.~\ref{fig:phonon}.
It is straightforward to generalize this result to the case of
finite temperature, see Appendix \ref{phonon_finiteT};
the error for $T = 50$ mK is shown as the dashed line in 
Fig.~\ref{fig:phonon}, showing almost identical behavior
to the zero-temperature case.

Our above analysis relied on a specific choice of
material (InAs) and device structure (3D heterostructure).
We expect that the results change qualitatively
for different setups.
A particularly relevant case, requiring a separate, 
detailed phonon analysis, is when the 1DTS is built
as a combination of a quasi-1D semiconducting nanowire
and an s-wave superconductor\cite{Lutchyn,Oreg,Mourik,Gharavi}. 
Since in these devices, the nanowire is supported by 
a substrate, and (at least partially) clamped by 
the superconductor and the normal-metal contact electrodes, 
the acoustic phonon modes might be gapped, 
resulting in very long charge relaxation times, 
potentially eliminating the phonon hazard from the 
parity readout procedure based on charge Rabi oscillations.

\section{Discussion}

\label{sec:discussion}

\subsection{Parity readout in a long Kitaev chain in the ideal limit}
\label{sec:longchain}

The problem we have studied so far was based on a minimal model, 
the two-site Kitaev chain, which might be realized in 
a double quantum dot proximitized by a superconductor\cite{LeijnsePoormans,SauChain,Fulga_2013}. 
To enable topologically protected qubits though, 
one needs to assemble longer chains, or 
consider proximitized quasi-one-dimensional nanostructures\cite{Alicea}.
Motivated by these observations, we now generalize
the two-site problem 
and here we consider a longer Kitaev chain, with $N>2$ sites, 
coupled to the readout dot, as shown in Fig.~\ref{fig:setup}c.

We still focus on the ideal Kitaev limit $v = \Delta$, but allow for disorder
of the on-site energies the same way as we did before
(see Eq.~\eqref{eq:twositekitaevchain} and 
section \ref{sec:slowchargenoise}). 
The  Hamiltonian is
$H = H_\text{chain} + H_\text{dot} + H_\text{tun}$, where
 \begin{subequations}
\bean
H_\text{chain} &=& 
\sum_{j=1}^N \varepsilon_j c_j^\dag c_j \\ \nonumber
&+&
\sum_{j=1}^{N-1}
v(c_j^\dag c_{j+1} + h.c.)
+
\Delta(c_j^\dag c_{j+1}^\dag + h.c.)
\\
H_\text{tun} &=& 
u(t) (c_1^\dag c_\text{dot} + c_N^\dag c_\text{dot} + h.c.),
\eean
\end{subequations}
and $H_\text{dot}$ is defined in Eq.~\eqref{eq:twositekitaevchain}.

The key question is: how do the error mechanisms studied
for the $N=2$ case change as we consider longer chains?
We  investigate the same error mechanisms: 
(A) leakage, 
(B) incomplete charge Rabi oscillation, and (C) charge relaxation. 

(A) The perturbative formula \eqref{eq:epsilontwosite} 
can be straightforwardly
extended to the $N>2$ case, 
yielding a result that differs only in a prefactor from 
the $N=2$ result of Eq.~\eqref{eq:epsilontwosite}:
\bean
\label{eq:epsilonNsite}
\epsilon = \frac{1}{8} \left(\frac{u_\textrm{max}}{\Delta}\right)^2.
\eean
Appendix \ref{app:leakage} contains the details of the 
derivation.
The energy level diagrams depicting the process triggered
by the readout tunnel pulse, for the $N=3$ and $N=4$ cases, 
are shown in Fig.~\ref{fig:levels}c-f.

(B) The analytical estimate of the parity readout error due to
slow charge noise and leakage,
in the limit $\sigma_\textrm{noise} \ll u_\textrm{max} \ll \Delta$, 
was given for $N=2$ in Eq.~\eqref{eq:twositeanalytical}.
This analytical result can be generalized to the $N>2$ case:
\bean
\label{eq:longchain}
\epsilon = \frac 1 8
\left(
 \frac{u_\text{max}}{\Delta}
 \right)^2 + \frac 1 4 
\left( \frac{\sigma_\textrm{noise}}{u_\text{max}}\right)^2
, \, \, \, (N>2).
\eean
That is, the error induced by the charge noise in the $N=2$ case is
different from the
$N>2$ cases, but in the latter cases it does not depend
on the chain length $N$. 

\begin{figure}
\centering
\includegraphics[width=1.0\columnwidth]{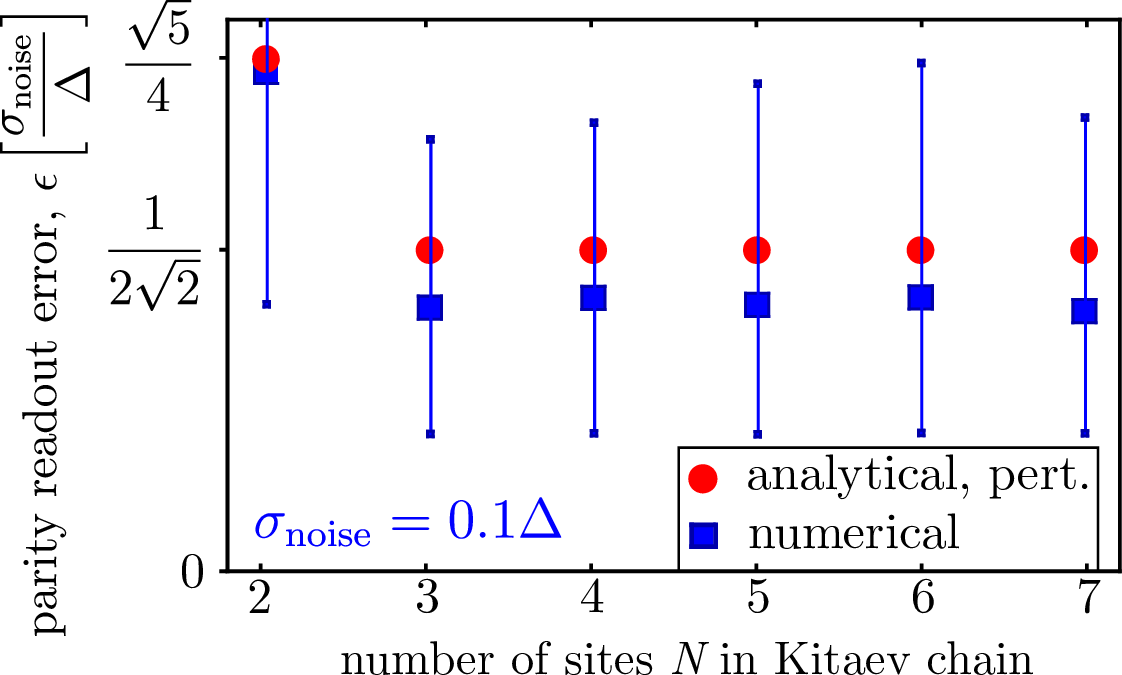}
\caption{
Length dependence of optimal leakage error.
Blue squares show the numerical results, 
red points show the perturbative analytical results
\eqref{eq:optimalerror2} (for $N=2$) 
and \eqref{eq:optimalerrorlong} (for $N>2$).
}
\label{fig:length}
\end{figure}

The error result \eqref{eq:longchain}, incorporating both 
leakage and incomplete Rabi oscillations, 
can be used to analytically estimate 
the optimal readout pulse strength and the corresponding minimal 
parity readout error.
The formula for the optimal readout pulse strength 
reads
\bean
\label{eq:optimalspeedlong}
u_\text{max}^\text{(opt)} = 2^{1/4} \sqrt{\sigma_\textrm{noise} \Delta},
\eean
and the corresponding minimal parity readout error 
is 
\bean
\label{eq:optimalerrorlong}
\epsilon^{\text{(min)}} = \frac{1}{2\sqrt{2}} \frac{\sigma_\textrm{noise}}{\Delta}.
\eean
The length dependence of the minimal parity readout error, 
according to this analytical estimate, is shown as 
the red points in Fig.~\ref{fig:length}.
The map of error thresholds corresponding to $N>2$ is shown in 
Fig.~\ref{fig:errorthresholds}b.

An important conclusion from Eq.~\eqref{eq:optimalerrorlong}
is that 
the optimal readout error as a function 
of chain length $N$ saturates already at 
$N=3$, and increasing the chain length does not bring
any further improvement. 
This saturation of the readout error for increasing system size
is in contrast to the length dependence of braiding-based
quantum gates: in the latter case, increasing the system size
can make the error exponentially suppressed\cite{Boross,Knapp_diabaticerror}.
This observation highlights a practical 
difficulty of the parity readout scheme studied here. 
It will be an interesting extension of this work to study how the
error in different parity readout schemes 
vary as the system size is increased, and whether there is a  
saturation effect similar to the one we have found. 
This task is particularly relevant for 
measurement-only topological quantum computing 
proposals\cite{Bonderson,KarzigScalable}, where
quantum gates are performed via sequences of measurements, instead
of braiding the anyonic defects.
%

To underpin the above analytical estimates, we have carried
out numerical simulations of the parity readout process. 
From these simulations, we have obtained the 
numerical values of the parity readout error shown in 
Fig.~\ref{fig:length} (blue squares), 
in good agreement with the analytical 
estimates (red points). 
The simulations follow the scheme 
described in section \ref{sec:slowchargenoise}. 
Error bars in Fig.~\ref{fig:length} connect
the 1st and 9th deciles of the $N_\text{r} = 5000$
different results corresponding to $N_\text{r}$ different
disorder realizations.
For these length-dependent simulations, 
we have used the Bogoliubov de
Gennes (BdG) formalism to compute
the time evolution, instead
of the direct Fock-space approach. 
We summarize the time-dependent BdG formalism in 
appendix \ref{BdG_appendix}. 
The advantage of the BdG formalism over the Fock-space
description is that the former requires diagonalizing
$2(N+1) \times 2(N+1)$ matrices, 
whereas the latter 
requires diagonalizing $2^{N+1} \times 2^{N+1}$ 
matrices.

(C) The length of the chain has no effect in our model 
of the phonon-mediated charge relaxation.
The reason is that the first-order correction of the energies
of the chain's even and odd ground states due to the
deformation potential is vanishing 
due to the usual topological protection, for any chain length
$N \geq 2$. 
This is exemplified by the Hamiltonians in Eqs.~\eqref{eq:hamiltonianeven}
and \eqref{eq:hamiltonianodd}, if $\varepsilon_1$, $\varepsilon_2$ and
$\varepsilon_\text{dot}$ are interpreted as deformation-induced
on-site energy shifts.
As a consequence, the model and results 
of phonon-mediated charge relaxation hold for longer chains $N>2$ 
as well. 
Similarly to the mechanisms (A) and (B), phonon-mediated
charge relaxation is not mitigated by lengthening chain.

\subsection{Smooth tunnel pulse leads to better readout}
\label{sec:diff_pulse}

So far, we have assumed that the tunnel pulse enabling
the parity-to-charge conversion has a square shape,
see Fig.~\ref{fig:setup}d and Eq.~(\ref{eq:squarepulse}).
It is natural to expect that the leakage error due
to the sudden turn-on and turn-off 
of such a
square pulse can be partially mitigated by using a smooth pulse
shape instead. 
Here, we present numerical results that
illustrate that this is indeed the case, 
and study how much is gained by using a smooth pulse. 

\begin{figure}
\centering
\includegraphics[width=1.0\columnwidth]{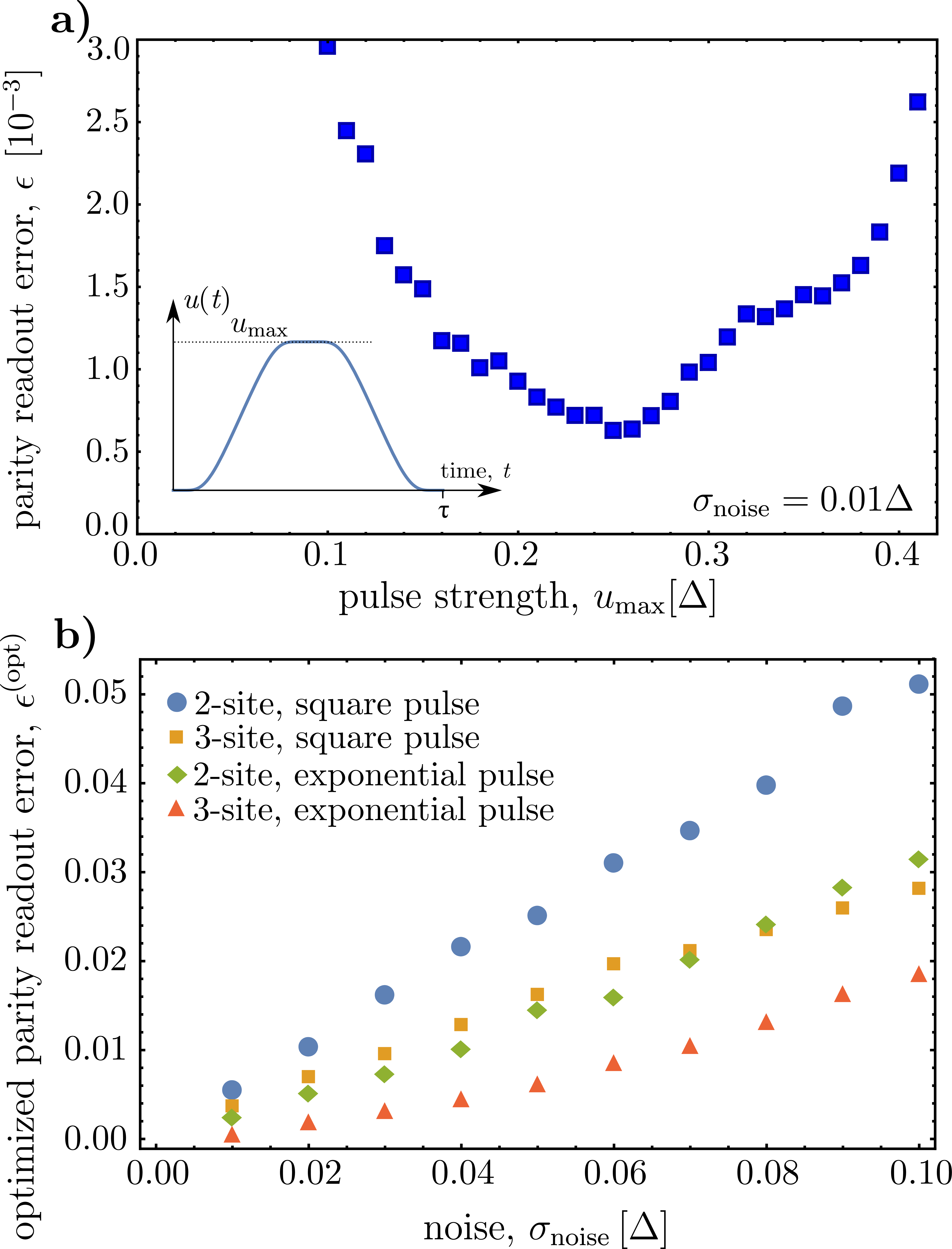}
\caption{Effect of the shape of tunnel pulse.
(a) Parity readout error averaged over numerically for 1000 disorder realizations (boxes) in case of  exponential-shape tunnel pulse (inset) and three-site Kitaev chain. (b) Optimized parity readout error as a function of noise strength for two- and three-site Kitaev chain in case of squared-shape and exponential-shape tunnel pulse. 
}
\label{fig:exppulse}
\end{figure}

In particular, here we use
a smooth, exponential time-dependent function,
as shown in the inset of Fig.~\ref{fig:exppulse}a:
\bean
u(t)=\left\{\begin{matrix} u_\textrm{max}\chi(2t/\tau),& \textrm{if } t<\tau/2  \\
 u_\textrm{max}\left(1-\chi(2t/\tau)\right),& \textrm{if } t\geq \tau/2,  \end{matrix}\right.
\eean
where the pulse shape function $\chi$ is defined as
\bean
\chi(x) = \frac{e^{-1/x}}{e^{-1/(1-x)}+e^{-1/x}}.
\eean
To realize a full Rabi oscillation,  the duration $\tau$
of the pulse is chosen to be 
$\tau=\frac{\pi\hbar}{u_\textrm{max}}$.
We solve the time-dependent Bogoliubov-de Gennes equation 
for this pulse shape numerically,
in a similar fashion as described above,
to calculate the disorder-averaged parity readout error
for various chain lengths.

In Fig.~\ref{fig:exppulse}a, we show the 
average parity readout error as a function of the tunnel pulse strength 
$u_\text{max}$, for a certain noise strength 
$\sigma_\textrm{noise}=0.01 \Delta$, 
in the case of a three-site Kitaev-chain. 
Comparing the results to those in Fig.~\ref{fig:noisy}a obtained for
the square pulse, we note the following similarities: 
(i) There is an optimal parity readout error at a finite pulse strength;
in Fig.~\ref{fig:exppulse}a it is $\approx 0.25 \, \Delta$.
(ii) Away from this optimal pulse strength, the error is increasing
for the same reasons as discussed in Sec.~\ref{sec:errors}. B and C.

One difference, however, is that the optimal parity readout error
with the smooth pulse (Fig.~\ref{fig:exppulse}a) is 
$\approx 5\cdot 10^{-4}$, which is approximately 
five times smaller than in the case of 
the square pulse. 
To illustrate the generality of this performance gain 
of the smooth pulse, 
we plot the optimal parity readout error as a function 
of noise strength $\sigma_\text{noise}$ in Fig.~\ref{fig:exppulse}b,
for two different chain lengths, comparing the 
performance of the square pulse and the smooth, exponential pulse. 
This plot shows that the using smooth pulse
improves the parity readout error  by a factor of $\sim 2$-$5$.  
We also observed that using different smooth pulses,
e.g., those with sine, sine-squared, etc., time dependence, 
we obtain almost the same optimized parity readout error as
with the exponential pulse. 

These results imply that the adiabatic errors of parity readout
in this setting can be significantly reduced by using
smooth tunnel pulses instead of  square pulses.

\subsection{Refinements of the noise model}

Our work provides an extension of earlier theory works
studying the effect of charge noise 
and phonons on readout of Majorana-based qubits.
Karzig et al.\cite{KarzigScalable} treated
charge noise phenomenologically
(see their Eqs.~(C21)-(C23)).
Similarly to Knapp et al.\cite{Knapp_dephasing}
(see their Eqs.~(43)-(44)),
both of these works studied noise effects that are beneficial for
readout. 
Munk et al.\cite{Munk_fidelity} provided a perturbative analysis
in equilibrium, which indicates that
the readout visibility and fidelity are influenced
by the entanglement of the device electrons and their
quantum-mechanical environment (see their section IV. B). 
Hyart et al.\cite{Hyart_majorana} describe a flux-controlled setup, 
and analyze the readout error due to photon shot noise of the readout 
cavity, 
a mechanism which is indeed practically relevant, although only
indirectly related to the physics of Majorana zero modes
(see their Appendix B). 
Munk et al. \cite {MunkReadout} and Steiner et al. \cite{Steiner}   investigated the effect of noise on Majorana readout induced by the measurement device.

We think that our quantitative analysis of
readout error due to leakage, charge noise and phonons,
which is based on 
transparent non-equilibrium minimal models possessing only the
key physical ingredients, will serve as  
important benchmarks for future works both in theory and experiment. 
Nevertheless, we envision that developing and understanding
a realistic parity-readout experiment will require a more elaborate
theoretical treatment, in which the 
electronic-structure models and 
the noise models applied 
in the present work are refined. 
For example, a distinct type of readout error might be 
caused by the high-frequency component of charge 
noise\cite{Dzurak,Cywinski,Mishmash}, 
which was neglected in our present work. 
Regardless of the noise model used, as long as noise is classical, 
the way to calculate readout error is similar to the way we used here: 
due to noise, the charge of the readout dot becomes a
random variable, and the task is to connect the statistical
characteristics of the noise to the probability distribution 
of the measured charge. 
Since the charge here is a binary variable, i.e., its measured
value is either 0 or 1, its distribution is completely characterized by its 
average, which we calculated in the present work for the
special case of quasistatic charge noise.

\subsection{Potential extensions of the phonon analysis}

In the phonon analysis of section \ref{sec:phonon}, 
we considered the deformation-potential electron-phonon 
interaction mechanisms. 
However, theoretical\cite{Khaetskii}
and experimental \cite{Elzerman2004ER,Hanson} 
results indicate that
the piezoelectric electron-phonon interaction is also important,
and can even dominate, in semiconductors without inversion symmetry,
such as the materials InAs and InSb frequently used in 
engineered topological superconductors\cite{Mourik,Laroche}.
Evaluating the effect of piezoelectricity on the readout
procedure studied here is
therefore an important future task. 
In principle, piezoelectric effects could be avoided by using 
host materials having inversion symmetry, 
such as silicon or germanium.

A further interesting extension of the present work could
aim at a more balanced description of the electron-phonon interaction. 
The key observation is that in our model,
phonons play two different roles, and these two roles
are represented as independent ingredients 
in the model, even though they arise from the same 
microscopic origin. 
First, the e-ph coupling provides
the glue of the Cooper pairs, and hence the e-ph coupling is hidden
in the effective superconducting gap $\Delta$. 
Second, the e-ph coupling contributes to the charge relaxation process 
leading to parity readout error.
Treating these two roles of the e-ph coupling on an equal footing would be 
an elegant and desirable extension of our work. 
We speculate that 
the semi-phenomenological model we used in this work might even
gain a microscopic foundation, e.g., in the case of 
superconductor-semiconductor heterostructures, 
where the gap could mostly be determined by the phonons
in the superconductor, whereas the relaxation 
could be dominated by the semiconductor phonons.

\subsection{Parity readout as a generalized measurement}

In this work, we characterized the parity readout error in 
terms of a single number between $0$ and $1$. 
A more thorough description of the readout procedure
can be given using generalized POVM measurements \cite{bookDiosi}.
The future extension of this work that includes this
refinement can provide a useful tool in post-processing the measured
data in quantum-information experiments.

\section{Conclusions}
\label{sec:conclusions}

In conclusion, we have studied error mechanisms affecting
the readout of a Majorana qubit in a one-dimensional
topological superconductor.
We describe (i) leakage due to a strong readout tunnel pulse, 
(ii) incomplete charge Rabi oscillations due to 
slow charge noise, and (iii) charge relaxation 
due to phonon emission and absorption. 
Using minimal models based on the Kitaev chain, 
we calculate the  readout error for these mechanisms.

Our work highlights and quantifies a practical difficulty with the 
Rabi-oscillation-based parity readout scheme we studied here. 
Namely, the readout process 
does not enjoy the same topological protection as the braiding-based
quantum gates, i.e., 
the readout error does saturates to a nonzero value as
the chain length increases.
This issue requires careful consideration, if this measurement scheme
is to be used in practical quantum-information experiments, 
especially for measurement-only quantum computing.

We also provide guidelines for experiments: given a certain 
level of charge noise and a certain phonon
environment, what are the requirements to perform 
a readout experiment with a certain small readout error?
For an example case where the device is 
based on a three-dimensional InAs nanostructure, 
we estimate that the error is caused mainly by 
leakage and slow charge noise, and phonon effects
are negligible. 
For this case, we provide analytical formulas on optimal
readout errors and how to achieve those. 
We expect that these results contribute 
to the fundamental understanding of topological 
qubit readout, and provide practical guidance for
near-term experiments aiming at topological qubit
demonstrations.

\acknowledgments

We acknowledge useful discussions with
J. Asb\'oth, M.-S. Choi, V. Derakhshan, J. K\"onig, R. Lutchyn, 
and G. Steffensen.
This research was supported by the National Research Development
and Innovation Office of Hungary within the 
Quantum Technology National Excellence Program (Project No. 
2017-1.2.1-NKP-2017-00001), grant FK124723 and BME-Nanotechnology FIKP grant (BME FIKP-NAT).
This work was completed in the ELTE Institutional Excellence Program
(1783-3/2018/FEKUTSRAT) supported
by the Hungarian Ministry of Human Capacities.

\appendix

\section{Low-energy description of parity-to-charge
conversion}
\label{app:lowenergy}

In section \ref{sec:paritytocharge}, 
we described the parity-to-charge conversion
mechanism in terms of the complete Fock-space
Hamiltonian. 
Here, we provide an insightful, approximate low-energy 
description\cite{FlensbergNonabelian}. 
Even though the low-energy nature of this approximation 
implies that it cannot directly be used to describe
the leakage error, it is useful to describe the charge-noise induced
and phonon-induced errors. 

The short informal summary of this low-energy 
picture, exemplified using the two-site Kitaev chain, 
is as follows. 
The ideal Kitaev chain hosts two zero-energy Majorana modes, 
which can be combined into a single zero-energy fermionic mode, 
such that the occupation of this mode (zero or one)
corresponds to the ground-state fermionic parity (odd or even). 
On the one hand, if this fermionic mode is empty in the
initial state (for $N=2$, this is the odd state), then this mode 
as well as the readout dot will remain empty at the end of the readout
procedure. 
On the other hand, if this fermionic mode is occupied in the inital
state (for $N=2$, this is the even state), 
then this excitation will be transferred to the dot by 
the end of the readout tunnel pulse, leaving the fermionic
mode empty but making the dot occupied. 

To formalize this, we rewrite the terms of the chain
Hamiltonian $H_\text{chain}$ 
and the chain-dot tunneling Hamiltonian $H_\text{tun}$
in terms of the Majorana operators
\bean
\gamma_{1A} &=& c^\dag_1 + c_1, \\
\gamma_{1B} &=& i(c_1 - c^\dag_1),
\eean
with analogous definitions for site 2.
In this Majorana representation, the Hamiltonians 
read
\bean
H_\text{chain} &=&
2 \Delta \gamma_{1B} \gamma_{2A},
\\
H_\text{tun} &=& \frac{u(t) }{2}
\left[
(\gamma_{1A} + i \gamma_{1B}) c_\text{dot}
+
(\gamma_{2A} + i \gamma_{2B}) c_\text{dot} + h.c. 
\right].
\eean
One observation is that $\gamma_{1A}$ and $\gamma_{2B}$ are
absent from $H_\text{chain}$. 
This implies that the nonlocal zero-energy mode 
\bean
d^\dag = \frac 1 2 \left( \gamma_{1A} + i \gamma_{2B} \right).
\eean
can be considered a low-energy eigenmode
of $H_\text{chain}$, and any further
Majorana operators are related to high-energy excitations and 
hence can be neglected in a low-energy approximation as
long as $u(t) \ll \Delta$. 
With this reasoning, we truncate the tunneling Hamiltonian 
to low energies, yielding the following approximation:
\bean
\label{eq:lowenergytunneling}
H_\text{tun} \approx u(t) \left(
d^\dag c_\text{dot} + h.c.
\right).
\eean
The final expression in Eq.~\eqref{eq:lowenergytunneling} 
reveals the simple picture anticipated in the previous paragraph. 

It is straightforward to generalize these considerations 
for any 1DTS, by using the actual Majorana zero
modes instead of 
$\gamma_{1A}$ and $\gamma_{2B}$. 
However, in such a generalized scenario, the electron transfer
probability can be finite for both the even and odd initial states. 
To illustrate this, let us generalize our previous 
calculations by modifying the tunnelling Hamiltonian 
to include two independent, complex-valued tunneling amplitudes:
\bean
H'_\text{tun} &=& 
\left(
	u_1 c^\dag_1 c_\text{dot} + h.c.
\right)
+
\left(
	u_2 c^\dag_2 c_\text{dot} + h.c.
\right)
\eean
Expressing this generalized tunneling Hamiltonian at low energies, 
in terms of the zero-energy fermionic mode $d^\dag$, 
we find
\bean
H'_\text{tun} \approx 
\frac{u_1 + u_2}{2} 
d^\dag c_\text{dot} 
+ 
\frac{u_1 - u_2}{2} 
d c_\text{dot} 
+ h.c.
\eean
This reveals that fine tuning $u_1 = u_2$ is required to 
completely forbid tunneling for the odd initial state. 

How is this fine-tuning requirement relevant for experiments? 
Let us exemplify this via the 1DTS system based on a semiconductor
nanowire proximitized by an $s$-wave superconductor\cite{Oreg,Lutchyn}. 
Here, a finite magnetic field is required to induce the 1DTS state. 
The presence of this magnetic field implies that time-reversal symmetry
is broken, so in general, tunnel amplitudes in a loop topology 
(see, e.g., Fig.~\ref{fig:setup}b,c) are complex-valued. 
In the setting considered here, where the tunnel amplitudes $u_1$ and
$u_2$ assist the tunneling process, a magnetic flux piercing the 
loop formed by the wire and the dot should be fine-tuned such that 
the complex phases of $u_1$ and $u_2$ are the same.
Of course, the absence of this fine tuning is not detrimental for 
the readout process, since the effective fermionic tunneling 
amplitude $(u_1 - u_2)/2$ for the odd state is in general
different from the amplitude $(u_1+u_2)/2$ for the even state;
fine tuning is required only to optimize the contrast between the
two initial states.

\section{Perturbative calculation for the leakage-induced 
parity readout error}
\label{app:leakage}

\subsection{Two-site Kitaev chain}
Here, we calculate the leakage-induced parity readout error in a 
two-site Kitaev chain (ideal Kitaev limit, $v = \Delta$)
during the readout protocol described in section \ref{sec:paritytocharge}.
We are focusing only the perturbative limit, when the strength of the readout tunnel pulse $u_\textrm{max}$ is much less than the 
effective superconducting gap 
of the Kitaev chain, i.e. $u_\textrm{max}\ll \Delta$. 
We will show that the parity readout error is well approximated by the simple result in Eq.~\eqref{eq:epsilontwosite}.

Without the tunnel pulse, the chain is uncoupled from the dot, and it has 
four energy eigenstates:
$\ket{\textrm{e}}$ and $\ket{\textrm{o}}$ at energy $-\Delta$, 
and $\ket{\textrm{e}'}$ and $\ket{\textrm{o}'}$ at $\Delta$.
Since the total parity is conserved during the readout process, 
we will separately investigate the cases where the parity 
is even or odd. 

First, we consider the odd subspace, see Eq.~\eqref{eq:hamiltonianodd}.
Here, the Hamiltonian $H_0 = H_\text{chain} + H_\text{dot}$ 
of the disjoint system, up to 
a global energy shift, reads
\bean \label{eq:Hodd}
H_0^{\textrm{odd}} &=& 2 \Delta 
\left(\ket{\textrm{e}',1}\bra{\textrm{e}',1} +  \ket{\textrm{o}',0}\bra{\textrm{o}',0}\right),
\eean
whereas the tunneling Hamiltonian is 
\bean
H_\text{tun}^{\textrm{odd}} &=& u_\textrm{max}\left(  \ket{\textrm{o}',0}\bra{\textrm{e}',1} - \ket{\textrm{e},1}\bra{\textrm{o}',0} + h.c.\right).
\eean
From the latter we see that the initial state $\ket{\textrm{o},0}$ 
does not couple to other states, hence the readout dot
remains empty during the readout, $P_{1 \leftarrow\textrm{o}}(t)=0$. 
This is also depicted in the level diagram
in Fig.\ref{fig:levels}b.

Second, we consider the even subspace.
Here, the Hamiltonian 
$H_0 = H_\text{chain} + H_\text{dot}$
of the disjoint system,
see Eq.~\eqref{eq:hamiltonianeven}, 
up to a global energy shift, reads
\bean
\label{eq:Heven}
H_0^{\textrm{even}} = 2\Delta \left( \ket{\textrm{e}',0}\bra{\textrm{e}',0} +  \ket{\textrm{o}',1}\bra{\textrm{o}',1}\right),
\eean
and the tunneling Hamiltonian is
\bean
H_\text{tun}^{\textrm{even}} = u_\textrm{max}\left( \ket{\textrm{o},1}\bra{\textrm{e},0} +  \ket{\textrm{o},1}\bra{\textrm{e}',0} + h.c.\right). 
\eean
The first term and its h.c. induces the charge 
Rabi oscillations allowing for the parity-to-charge
conversion. 
The second term and its h.c. induces leakage from the 
relevant subspace, and this leakage leads to 
imperfect readout.

Treating $H^\text{even}_\text{tun}$ as the perturbation, 
we use perturbation theory to obtain the
eigenstates $\ket{j}$ 
($j\in \{1,2,3,4\}$) of $H^\text{even} = 
H^\text{even}_0 + H^\text{even}_\text{tun}$
and the energies $E_j$
up to second order in $u_\textrm{max}/\Delta$. 
If the inital state is $\ket{\textrm{e},0}$, than the probability that the QD is occupied after a time $t$ is
\bean \label{eq:prob}
P_{1\leftarrow \textrm{e}}(t) = 
\left|\bra{\textrm{o},1}U(t)\ket{\textrm{e},0}\right|^2
+
\left|\bra{\textrm{o}',1}U(t)\ket{\textrm{e},0}\right|^2,
\eean
where the time-evolution operator is $U(t)=e^{-iE_jt/\hbar}\ket{j}\bra{j}$.  
In Eq.~\eqref{eq:prob}, the second term vanishes because
the state 
$\ket{\textrm{o}',1}$ is not populated
 during the time evolution of $\ket{\textrm{e},0}$. 
To evaluate $P_{1 \leftarrow \textrm{e}}$ in Eq.~\eqref{eq:prob},
we drop those terms of the projectors $\ket{j}\bra{j}$ in 
the propagator $U(t)$ that are beyond second order in
$u_\textrm{max}/\Delta$, yielding
\bean 
\label{eq:perturbativeshortchain}
P_{1\leftarrow \textrm{e}}(t) \approx
\left[1-\frac{5}{16}\left(\frac{u_\textrm{max}}{\Delta}\right)^2\right]
\sin^2 \left(\frac{u_\textrm{max}}{\hbar} t\right).
\eean
This  expression reaches its maximum at $t=\frac{\pi\hbar}{2 u_\textrm{max}}$.
Hence,  the leakage does not change the optimal 
duration of the tunnel pulse. 
This maximum of $P_{1 \leftarrow \textrm{e}}(t)$ is below 1, hence
the readout is imperfect, 
and the corresponding parity readout error can be read off
from Eq.~\eqref{eq:perturbativeshortchain}, and is given
by Eq.~\eqref{eq:epsilontwosite}.

\subsection{Three-site or longer Kitaev chains}

In a three-site or longer Kitaev chain, there are more than 
two high-energy states, see Fig.~\ref{fig:levels}c-f.
However, in the ideal Kitaev limit studied here, 
only two of them, $\ket{\textrm{o}'}$ and $\ket{\textrm{e}'}$, are coupled to 
the low energy ground states $\ket{\textrm{o}}$ and $\ket{\textrm{e}}$,
by the readout pulse. We restrict our calculation to these four 
states and show that the leakage-induced parity readout
error for $N>2$ is given by Eq.~\eqref{eq:epsilonNsite}.

In case of a three-site chain in the ideal Kitaev limit, 
the initial state $\ket{\textrm{e},0}$ remains unchanged (see Fig.~\ref{fig:levels}c), 
whereas 
the initial state $\ket{\textrm{o},0}$ evolves into $\ket{\textrm{e},1}$ due to the readout pulse (see Fig.~\ref{fig:levels}d).
The coupling to the high-energy states makes readout imperfect, 
and leads to a nonzero parity readout error.
The unperturbed Hamiltonian in the odd subspace is the same as in Eq. (\ref{eq:Hodd}), but the perturbation has a different form, 
namely
\bean 
H_1^{\textrm{odd}} &=& u_\textrm{max}\left( \ket{\textrm{e},1}\bra{\textrm{o},0} + h.c.\right)+\nonumber\\
&&\frac{u_\textrm{max}}{\sqrt{2}}\left( \ket{\textrm{e},1}\bra{\textrm{o}',0} + \ket{\textrm{o},0}\bra{\textrm{e}',1} + h.c.\right).
\eean
Following the method described in the previous subsection, we can calculate the dot occupation at time $t$, 
and find
\bean \label{eq:prob2}
P_{1 \leftarrow \text{o}}(t) &= &
\left|\bra{\textrm{e},1}U(t)\ket{\textrm{o},0}\right|^2 + \left|\bra{\textrm{e}',1}U(t)\ket{\textrm{o},0}\right|^2 \\ 
&=&\sin^2\left(\frac{tu_\textrm{max}}{\hbar}\right)-\frac{u_\textrm{max}^2}{8\Delta^2}+\frac{u_\textrm{max}^2}{16\Delta^2}\cos\left(\frac{tu_\textrm{max}}{\hbar}\right)\nonumber
\\
&\times &\left\{6\cos\left(\frac{tu_\textrm{max}}{\hbar}\right)-4\cos\left[\frac{t}{2\hbar\Delta}(4\Delta^2+u_\textrm{max}^2)\right]\right\}.
\nonumber
\eean
This function has a maximum at $t=\frac{\pi\hbar}{2u_\textrm{max}}+\mathcal{O}\left(\frac{u_\textrm{max}\hbar}{\Delta^2}\right)$, so the optimal length of the pulse gets a small correction, compared 
to the leakage-free case. The leading-order value $\frac{u_\textrm{max}^2}{8\Delta^2}$ of the parity readout error is coming from the maximal value of Eq. \ref{eq:prob2}.

In the even subspace, the unperturbed Hamiltonian is that in 
Eq. (\ref{eq:Heven}), but the perturbation is
\bean 
H_\text{tun}^{\textrm{even}} = 
u_\textrm{max}\left(\frac{1}{\sqrt{2}} \ket{\textrm{e},0}\bra{\textrm{o}',1} + \ket{\textrm{o}',1}\bra{\textrm{e}',0} + h.c.\right).
\eean
The readout tunnel pulse gives no matrix element between the ground states, therefore the initial state $\ket{\textrm{e},0}$ is stationary
if we disregard higher-energy states. 
In contrast to the 2-site case, 
the state $\ket{\textrm{e},0}$ is not decoupled from the other states, 
but is coupled to high-energy states.  
This feature provides a small but finite probability for the dot
being occupied at the end of the tunnel pulse:
\bean 
P_{1 \leftarrow \text{e}}(t= \pi/(2u_\text{max}))
 = \frac{u_\textrm{max}^2}{8\Delta^2}.
\eean
Hence, we conclude that the parity readout error is 
indeed given by Eq.~\eqref{eq:epsilonNsite} in this case. 

We can generalize this result for $N>3$. 
Whenever $N$ is odd, then the leading-order dynamics is
that the initial state $\ket{\textrm{e},0}$  
evolves into the state $\ket{\textrm{o},1}$ 
and hence the charge of the dot changes from 0 to 1,
whereas the initial state $\ket{\textrm{o},0}$ remains unchanged. 
Coupling to the high-energy states spoils the perfect parity-to-charge 
conversion.  
The level diagrams of the most relevant states 
are analogous to those in Figs.~\ref{fig:levels}c and d, and this
implies that the parity readout error in the perturbative limit is
given by Eq.~\eqref{eq:epsilonNsite}. Furthermore, the same conclusion
can be derived for the cases when $N$ is even and $N>3$.

\section{Perturbative calculation of the parity readout
error due to slow charge noise}
\label{app:slowchargenoise}

In the main text, we have provided analytical perturbative
results, Eqs.~\eqref{eq:twositeanalytical} and \eqref{eq:longchain} 
for the parity readout 
error in the presence of slow charge noise. 
Here we provide the details leading to those results. 

As discussed in the main text, we consider uncorrelated quasi-static 
on-site energy noise (disorder) on the chain sites as well as 
on the dot, and represent disorder with Gaussian random 
variables with standard deviation $\sigma_\textrm{noise}$.
We consider a specific perturbative limit, defined
by $\sigma_\textrm{noise}\ll u_\textrm{max}\ll \Delta$.
We will show that the value of noise-induced parity readout error is 
$\frac{\sigma_\textrm{noise}^2}{4u_\textrm{max}^2}$, irrespective
of the chain length $N$.

Recall that the 
on-site energies of the QDs in the Kitaev chain are denoted by 
$\varepsilon_i$, where $i\in\{1, \cdots, N\}$, furthermore, the on-site energy 
of the readout dot is denoted by $\varepsilon_\text{dot}$. 
The Hamiltonian of the
globally even sector,
 projected onto the $2\times2$ subspace spanned by the ground states $\ket{\textrm{e},0}$ and $\ket{\textrm{o},1}$:
\bean 
H=\varepsilon_\text{dot}\ket{\textrm{o},1}\bra{\textrm{o},1} + u_\textrm{max} \left(\ket{\textrm{e},0}\bra{\textrm{o},1}+h.c.\right),
\eean
where only the onsite-energy of the dot appears. 
Note that for simplicity, we use a notation corresponding to an
even chain length $N$, but all conclusions are also valid
for an odd $N$ by the replacements $\text{e} \leftrightarrow \text{o}$.

If the initial state is $\ket{\textrm{e},0}$, then 
the dot occupation probability 
after a time $t$ is
\bean 
P_{1 \leftarrow \text{e}}(t)=\frac{4u_\textrm{max}^2}{4u_\textrm{max}^2+\varepsilon_\text{dot}^2}\sin^2{\left(\frac{1}{2\hbar}t\sqrt{4u_\textrm{max}^2+\varepsilon_\text{dot}^2}\right)},
\eean
and $P_{0 \leftarrow \text{e}}(t) = 1- P_{1\leftarrow \text{e}}$.
Taking the disorder-averaged value 
$\bar{P}_{0 \leftarrow \text{e}}$ of the 
probability $P_{0 \leftarrow \text{e}}(t)$ at time
$t = \pi \hbar / (2 u_\text{max})$, 
we find 
\bean 
\bar{P}_{0 \leftarrow \text{e}} &=&
\int_{-\infty}^\infty d\varepsilon_\text{dot} P_{0 \leftarrow \text{e}}\left(t=\frac{\pi\hbar}{2u_\textrm{max}}\right) \frac{e^{-\frac{\varepsilon_\text{dot}^2}{2\sigma_\textrm{noise}^2}}}{\sqrt{2\pi}\sigma_\textrm{noise}} \nonumber\\
&&\approx \frac{\sigma_\textrm{noise}^2}{4u_\textrm{max}^2}.
\eean
In the last approximation, we applied the hierarchy of the parameters $\sigma_\textrm{noise}\ll u_\textrm{max}\ll \Delta$.

\section{Charge relaxation due to phonon emission at finite temperature}

\label{phonon_finiteT}

In section \ref{sec:phonon} of the main text, we presented
results for the parity readout error due to phonon-mediated
inelastic processes. There, we focused on the zero-temperature
case. 
Here, we provide details of the calculation of the finite-temperature
case.
Compared to the zero temperature limit where only 
spontaneous emission of phonons influence the parity-to-charge
conversion, here we discuss the finite-temperature effects
of stimulated emission and absorption of phonons as well.

First, we generalize Eq. (\ref{eq:phononerror2}) to 
derive the relation  
between the parity readout error $\epsilon$ and the inelastic 
rates.
In this case, we have an energetically downhill
rate $\Gamma_\downarrow$, accounting for
phonon emission and de-excitation from state $\ket{a}$ to 
state $\ket{b}$, and an uphill rate $\Gamma_\uparrow$, 
accounting for phonon absorption and excitation from 
$\ket{b}$ to $\ket{a}$.
The time evolution of the density matrix is described by the 
Bloch-Redfield equation\cite{Makhlin}:
\bean
\dot{\rho}_\textrm{bb}(t) = -\Gamma_\uparrow \rho_\textrm{bb}(t) + \Gamma_\downarrow \rho_\textrm{aa}(t), \\
\dot{\rho}_\textrm{aa} (t)= \Gamma_\uparrow \rho_\textrm{bb}(t) - \Gamma_\downarrow \rho_\textrm{aa}(t), \\
\dot{\rho}_\textrm{ab} (t)= \frac{2iu_\textrm{max}}{\hbar} {\rho}_\textrm{ab}(t) -\frac{1}{T_2}{\rho}_\textrm{ab}(t),
\eean
where $\frac{1}{T_2}=\frac{\Gamma_\uparrow + \Gamma_\downarrow}{2}$. 
Initially, our state is  $\ket{\textrm{e},0}$ , 
so $\rho_\textrm{bb}(0)=\rho_\textrm{aa}(0)=\rho_\textrm{ab}(0)=\rho_\textrm{ba}(0)=\frac{1}{2}$. The parity readout error $\epsilon$ is the probability of finding the readout dot empty  after a time $t=\tau_\textrm{ideal}=\frac{\hbar \pi}{2u_\textrm{max}}$, that is,
\bean 
\label{eq:finitetemperatureerror}
\epsilon= \frac{{\rho}_\textrm{bb}(\tau_\text{ideal}) + {\rho}_\textrm{ab}(\tau_\text{ideal}) + {\rho}_\textrm{ba}(\tau_\text{ideal}) + {\rho}_\textrm{aa}(\tau_\text{ideal})}{2}.
\eean
The trace of the density matrix does not change in time, 
i.e., ${\rho}_\textrm{bb}(t)+ {\rho}_\textrm{aa}(t)=1$. 
Solving the differential equation of $\rho_\textrm{ba}(t)$, we obtain 
\bean \label{Pt}
\rho_\textrm{ba}(t) = \frac{1}{2}e^{\frac{2iu_\textrm{max}t}{\hbar}-\frac{t}{T_2}}.
\eean
Then, we substitute this solution to Eq.~\eqref{eq:finitetemperatureerror},
insert $t = \tau_\text{ideal}$, and thereby obtain,
as a generalization of Eq.~\eqref{eq:phononerror2}
\bean 
\label{eq:epsilonX}
\epsilon &=&\frac{1}{2} \left[1-e^{-\frac{\tau_\textrm{ideal}}{T_2}}\right] \approx \frac{\tau_\textrm{ideal}}{2T_2} =\frac{\hbar\pi\left(\Gamma_\uparrow + \Gamma_\downarrow\right)}{8u_\textrm{max}}.
\eean

We also need the finite-temperature values of the
transition rates. As usual, the rate of stimulated
emission and the rate of absorption are both equal to 
the rate of spontaneous emission multiplied
by the Bose-Einstein function at the transition energy,
that is, 
$n_\text{BE}(2 u_\text{max},T) = 
\left(e^{\frac{2u_\text{max}}{k_\text{B} T}}-1\right)^{-1}$.
In formulas, this means
\bean
\Gamma_\downarrow(T) &=& \Gamma_\downarrow(T = 0) 
\left[1+n_\text{BE}(2 u_\text{max},T)\right],\\
\Gamma_\uparrow(T) &=& 
\Gamma_\downarrow(T = 0) \, n_\text{BE}(2 u_\text{max},T).
\eean
Combining these expressions with Eq.~\eqref{eq:epsilonX},
we obtain the parity readout error as \bean 
\epsilon &=& \frac{1}{4}T_\textrm{Rabi}\left(\Gamma_\uparrow + \Gamma_\downarrow\right)  = \frac{\Xi^2 u_\textrm{max}^2}{8\rho\hbar^3c_\textrm{L}^5} \coth{\left(\frac{u_\textrm{max}}{k_BT}\right)}.
\eean
For a 3D bulk InAs device,
see material parameters in the main text, this result reads as
\bean
\epsilon =
\left(
 \frac{u_\textrm{max}}{2.61 \textrm{meV}}
 \right)^2
 \coth{\left(\frac{u_\textrm{max}}{k_\text{B} T}\right)}.
\eean
In Fig. \ref{fig:phonon}, we compare the phonon-induced
parity readout error at zero temperature (solid black) and at 
$T= 50 \;\textrm{mK}$ (dashed gray),
corresponding to the energy scale
$k_\text{B} T \approx 4.3\;\mu\textrm{eV}$.

\section{Bogoliubov-de Gennes formalism for
dynamics} \label{BdG_appendix}

In section \ref{sec:longchain},
we have considered the dependence of the parity readout error $\epsilon$
on the chain length $N$. 
The numerical results in Fig.~\ref{fig:length} are obtained by 
applying the Bogoliubov-de Gennes (BdG) formalism to describe
the dynamics\cite{Scheurer} of the system. 
Here, we outline this formalism. 
For concreteness, we describe the special case when the readout
dot is coupled to the two-site Kitaev chain, 
$N=2$; generalization is straightforward. 

The setup is modelled by the 
time-dependent Hamiltonian 
$H(t)$ in 
Eq.~\eqref{eq:twositekitaevchain}. 
Once we define the column vector 
$\tilde{\mathbf{c}}$ of 
 the local fermionic operators via
\bean
\tilde{\mathbf{c}} = \left(\bna{c}
	c_1 \\
	c_2 \\
	c_\text{dot} \\
	c^\dag_1 \\
	c^\dag_2 \\
	c^\dag_\text{dot}
\eda \right),
\eean
we can rearrange the Hamiltonian $H$ as
\bean
H(t)=\frac{1}{2}\tilde{\mathbf{c}}^\dag \mathcal{H}(t) 
\tilde{\mathbf{c}}
+ \frac{\epsilon_1 + \epsilon_2 + \epsilon_\textrm{dot}}{2}.
\eean
Here, we have introduced the BdG matrix via
\bean
\mathcal{H}(t)=\begin{pmatrix}
    \epsilon_1  & v & u(t)&0 &\Delta &0 \\
   v & \epsilon_2  & u(t) & -\Delta& 0& 0\\
		u(t) & u(t) &\epsilon_\textrm{dot}  &0 &0 &0 \\
		0 & -\Delta &0 &-\epsilon_1 & -v & -u(t)\\
		\Delta &0 &0 & -v& -\epsilon_2 & -u(t)\\
		0 & 0 &0 & -u(t)& -u(t)& -\epsilon_\textrm{dot}
\end{pmatrix},
\eean
and the dagger in  $\tilde{\mathbf{c}}^\dag$ implies 
element-wise hermitian conjugation as well as transposing 
the column vector to transform it into a row vector. 

The form of $\mathcal{H}(t)$ is specified by the requirement
that it is particle-hole symmetric for all times $t$, that is, 
\bean
 \mathcal P \mathcal{H}(t) \mathcal P^{-1} = - \mathcal{H}(t),
\eean 
where $\mathcal P = \sigma_x K$ is the canonical particle-hole
transformation, with $\sigma_x$ being the first Pauli matrix
acting on the Nambu (particle-hole) degree of freedom, 
and $K$ is complex conjugation.

Recall that in the readout protocol, 
initially there is no tunneling between the Kitaev chain and 
the readout dot, $u(t=0)=0$. 
The corresponding Hamiltonian 
(BdG matrix)
will be denoted as
$H_i$ ($\mathcal{H}_i$).
When the tunnel pulse is switched on for $0<t<\tau$, 
then $u(t)=u_\textrm{max}$.
The corresponding Hamiltonian (BdG matrix) will be 
denoted as $H_f$ ($\mathcal{H}_f$).

Our goal is to describe the time evolution of the expectation value of
an observable, for two different initial states.
Our observable is the dot occupation
$n_\text{dot} = c^\dag_\text{dot} c_\text{dot}$,
since this
is used to characterize the parity readout error.
The two different initial states are the even and odd ground states
of the chain, with empty dots in both cases. 
To be able specify the initial states more precisely, we first
discuss the relation between the many-body energy eigenstates
and the BdG matrix. 

By solving the eigenvalue problem of the initial BdG matrix 
$\mathcal{H}_i$, we can find the six eigenvalues $\lambda_i$ 
and the corresponding eigenvectors (column vectors) $\phi_i$,
with $i =1,2,\dots, 6$.
Due to the particle-hole symmetry of the BdG matrix, 
it is possible (and convenient) to order the eigenvalues 
such that $\lambda_1, \lambda_2, \lambda_3 >0$
and $\lambda_4 = -\lambda_1$,
$\lambda_5 = -\lambda_2$,
$\lambda_6 = - \lambda_3$.
We can also choose the eigenvectors such that
$\phi_4 = \mathcal P \phi_1$,
$\phi_5 = \mathcal P \phi_2$,
$\phi_6 = \mathcal P \phi_3$.

Using the eigenvectors $\phi_i$, we can express the unitary diagonalizer 
of $\mathcal{H}_i$ via
\bean
U_\textrm{diag} = \left(\bna{c}
\phi^\dag_1 \\ \dots \\ \phi^\dag_6
\eda \right),
\eean
and the quasiparticle operators via
\bean
\tilde{\mathbf{d}} = U_\text{diag} \tilde{\mathbf{c}}.
\eean
The Hamiltonian $H_i$ is diagonalized, or in other words, 
it is written as the Hamiltonian of non-interacting quasiparticles, 
as
\bean
H_i =\frac{1}{2} \tilde{\mathbf{d}}^\dag 
\mathcal{H}_i^{(\text{d})}
\tilde{\mathbf{d}}+\frac{\epsilon_1 + \epsilon_2 + \epsilon_\textrm{dot}}{2}.
\eean 
with
\bean
\mathcal{H}_i^{(\text{d})} =
\text{diag}(\lambda_1,\lambda_2,\lambda_3, -\lambda_1, -\lambda_2, -\lambda_3).
\eean
Due to the particle-hole symmetry exploited above, 
we can label the elements of the 
vector $\tilde{\mathbf{d}}$ as
\bean
\tilde{\mathbf{d}} = \left( \bna{c}
d_1 \\ d_2 \\ d_3 \\ d_1^\dag \\ d_2^\dag \\ d_3^\dag
\eda \right).
\eean

Recall that our goal is to calculate the time dependence of the
dot occupation, e.g., 
\bean \label{ndot}
P_{1 \leftarrow e}(t) =  \bra{\Psi_\text{e}} \left[\tilde{\mathbf{c}}(t)\right]_6 \left[\tilde{\mathbf{c}}(t)\right]_3  \ket{\Psi_\text{e}}
\eean
where $[\ldots]_j$ is the $j$th component of the vector, 
$\ket{\Psi_\text{e}}$ denotes the even-parity initial state,
and the elements of the vector $\tilde{\mathbf{c}}(t)$ 
are the elements of the vector $\tilde{\mathbf{c}}$,
transformed to the Heisenberg picture:
\bean \label{dt}
\tilde{\mathbf{c}}(t)& =& e^{iH_ft/\hbar}\; \tilde{\mathbf{c}}\; e^{-iH_ft/\hbar}=\nonumber\\
&=& e^{i\frac{1}{2\hbar}\tilde{\mathbf{c}}^\dag \mathcal{H}_f 
\tilde{\mathbf{c}}t}   \; \tilde{\mathbf{c}} \; e^{-i\frac{1}{2\hbar}\tilde{\mathbf{c}}^\dag \mathcal{H}_f 
\tilde{\mathbf{c}}t}=\nonumber\\
&=& e^{-\frac{i}{\hbar}\mathcal{H}_ft}\;\tilde{\mathbf{c}}.
\label{eq:heisenberg}
\eean
For clarity, we remark that in the first and second lines of 
Eq.~\eqref{eq:heisenberg}, the exponentialized operators
are acting element-wise, on the Fock-space operators
forming the vector $\tilde{\mathbf{c}}$. 
However, in the third line, the exponentialized
operator is a $6 \times 6$ matrix acting on the 6-dimensional
vector formed by $\tilde{\mathbf{c}}$.

Inserting Eq.~\eqref{eq:heisenberg} into Eq. (\ref{ndot}) yields
\bean
P_{1 \leftarrow \text{e}}(t) &=& \bra{\Psi_\text{e}} \left[e^{-\frac{i}{\hbar}\mathcal{H}_ft} U_\textrm{diag}^\dag \tilde{\mathbf{d}}\right]_6 \left[e^{-\frac{i}{\hbar}\mathcal{H}_ft} U_\textrm{diag}^\dag \tilde{\mathbf{d}}\right]_3  \ket{\Psi_\text{e}}\nonumber\\
&\equiv &\sum_{m,n=1}^{6} S_{mn}(t) \bra{\Psi_\text{e}} \tilde{d}_m\tilde{d}_n \ket{\Psi_\text{e}}, \label{eq:p1e}
\eean
where the second line is 
an implicit 
definition of the matrix 
\bean
\label{eq:matrixs}
S_{mn}(t)=\left[e^{-\frac{i}{\hbar}\mathcal{H}_ft}U_\textrm{diag}^\dag\right]_{6m}\left[e^{-\frac{i}{\hbar}\mathcal{H}_ft} U_\textrm{diag}^\dag \right]_{3n}.
\eean
Following our notation introduced earlier, 
$\tilde{d}_m$ denotes the $m$th component of the vector 
$\mathbf{\tilde{d}}$.
Evaluating the matrix elements in Eq.~\eqref{eq:p1e} yields
\bean
\label{eq:p1eresult}
P_{1 \leftarrow \text{e}}(t)  =
\sum_{\mu = 1}^3 S_{\mu,3+\mu}(t).
\eean
This can be evaluated after the propagator and the
BdG eigenvectors in Eq.~\eqref{eq:matrixs} have been obtained, 
e.g., numerically.

So far, we have assumed that our initial state
is the even ground state $\ket{\Psi_\text{e}}$ 
of our initial Hamiltonian $H_i$. 
Depending on the values of the on-site energies $\varepsilon_1$
and $\varepsilon_2$, the actual ground state of the
wire can be either even or odd. 
For small on-site energy disorder $\sigma_\text{noise} \ll v = \Delta$, 
as considered in the main text, the lowest positive 
BdG quasiparticle excitation energy, say $\lambda_1$, is much smaller
than the other quasiparticle excitation energies.
If the even ground state is the actual ground state, 
then the odd ground state is 
$\ket{\Psi_\text{o}} = d^\dag_1 \ket{\Psi_\text{e}}$.
If the initial state is this odd ground state, 
then the time-dependent occupation probability of the dot 
reads 
\bean
P_{1 \leftarrow \text{o}}(t) = 
\sum_{m,n=1}^{6} S_{mn}(t) \braket{\Psi_\text{e} | d_1 \tilde{d}_m \tilde{d}_n 
d^\dag_1 | \Psi_\text{e} }.
\eean
By evaluating the matrix elements, we find
\bean
\label{eq:p1oresult}
P_{1 \leftarrow \text{o}}(t) = S_{4,1}(t) + \sum_{\mu = 2}^3 S_{\mu,3+\mu}(t).
\eean
We have used the expressions
\eqref{eq:matrixs}, \eqref{eq:p1eresult}, \eqref{eq:p1oresult} 
to perform the numerical calculations leading to Fig.~\ref{fig:length}.


\bibliography{ParityReadout}

\end{document}